\documentclass[aip,graphicx,reprint]{revtex4-1}

\draft 

\usepackage{graphicx}
\usepackage{dcolumn}
\usepackage{bm}

\usepackage[utf8]{inputenc}
\usepackage[T1]{fontenc}
\usepackage{mathptmx}
\usepackage{etoolbox}

\usepackage{amssymb, color}
\usepackage{hyperref}
\usepackage{natbib}
\usepackage{booktabs,tabulary}
\usepackage{multirow}
\usepackage{amsmath,soul}
\usepackage[export]{adjustbox}
\usepackage{float}
\usepackage[caption=false]{subfig}
\usepackage{array}
\usepackage{cases}
\usepackage{makecell}
\usepackage{physics}
\usepackage{yhmath}

\usepackage[colorinlistoftodos,shadow,textwidth=18mm]{todonotes}

\newcommand{\E}[1]{\left\langle#1\right\rangle}

\graphicspath{{../}}
\makeatletter
\def\@email#1#2{%
 \endgroup
 \patchcmd{\titleblock@produce}
  {\frontmatter@RRAPformat}
  {\frontmatter@RRAPformat{\produce@RRAP{*#1\href{mailto:#2}{#2}}}\frontmatter@RRAPformat}
  {}{}
}%
\makeatother
\begin{document}


\title{Geometric model of crack-templated networks for transparent conductive films}

\author{Jaeuk Kim}
\affiliation{McKetta Department of Chemical Engineering, University of Texas at Austin, Austin, Texas 78712, United States}

\author{Thomas M. Truskett}
\email{truskett@che.utexas.edu}
\affiliation{McKetta Department of Chemical Engineering, University of Texas at Austin, Austin, Texas 78712, United States}
\affiliation{Department of Physics, University of Texas at Austin, Austin, Texas 78712, United States}

\date{\today}

\begin{abstract}
Crack-templated networks, metallic frameworks fabricated from crack patterns in sacrificial thin films, can exhibit high optical transmittance, high electric conductivity, and a host of other properties attractive for applications. Despite advances in preparing, characterizing, and analyzing optoelectronic performance of cracked template networks, limited efforts have focused on predicting how their disordered structures help determine their electrical and optical properties and explain their interrelationships. We introduce a geometric modeling approach for crack-templated networks and use simulation to compute their wavelength- and incident angle-dependent optical transmittance and sheet resistivity. We explore how these properties relate to one another and to those of metallic meshes with periodically ordered aperture arrays. We consider implications of the results for optoelectronic applications, compare figure-of-merit predictions to experimental data, and highlight an opportunity to extend the modeling approach using inverse methods.
\end{abstract}


\maketitle 

Transparent conductive films, characterized by both high optical transmittance and high direct current (DC) electrical conductivity, have found wide application, including in flat panel displays and touch screens, smart windows and glass facades, as well as systems for harvesting solar energy.\cite{castellon2018transparent}
Films of transparent oxide conductors are well-suited and commonly adopted for many such technologies, though the search continues for replacements with higher flexibility and stretchability, higher transmittance in the IR and UV, and lower cost. Metallic networks represent one class of promising alternatives. We focus here on disordered metallic networks,\cite{han_uniform_2014,ye_metal_2014,gao_physics_2016, gao_nature-inspired_2018} which have advantages over ordered, lithographically patterned networks in terms of scalability,\cite{ye_metal_2014,jung2020recent,muzzillo2020patterning} while also avoiding undesirable interference-related iridescence,\cite{qiu_metallic_2016, wen_bright_2021,chen_non-iridescent_2021} Moir\'{e},\cite{suh2016random} or starburst\cite{jung2019moire} effects characteristic of periodically patterned grids or meshes. Geometric features of metallic networks, such as the cross-sections and connectivities of conductive paths and the size distributions of the intervening voids, play vital roles in determining their sheet resistance, light transmittance, and other measures of optoelectronic performance.

\begin{figure*}[ht!]
    \subfloat[]{\includegraphics[width=0.7\textwidth]{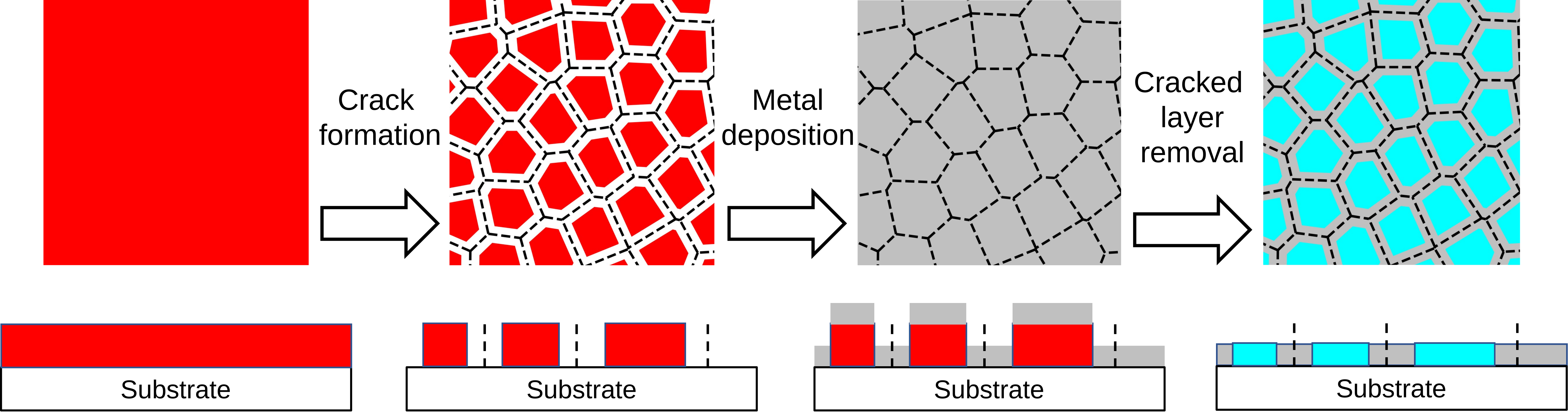}}
    
    \subfloat[]{\includegraphics[width=0.6\textwidth]{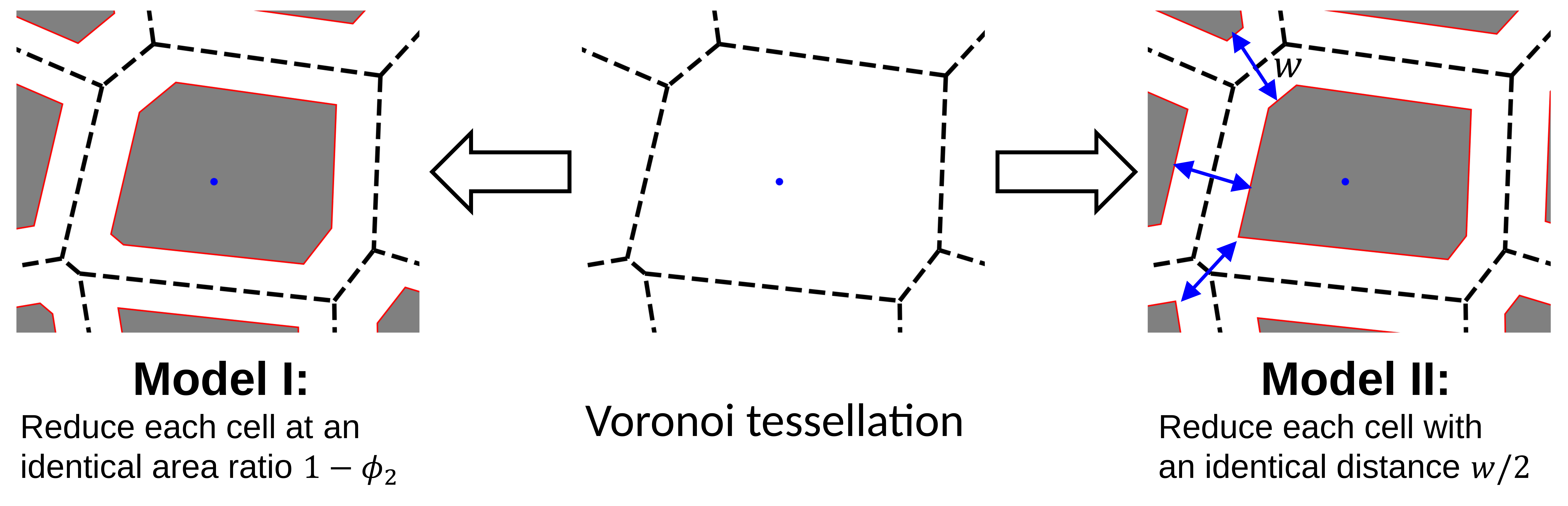}}
    \hspace{5pt}
    \subfloat[]{\includegraphics[width=0.35\textwidth]{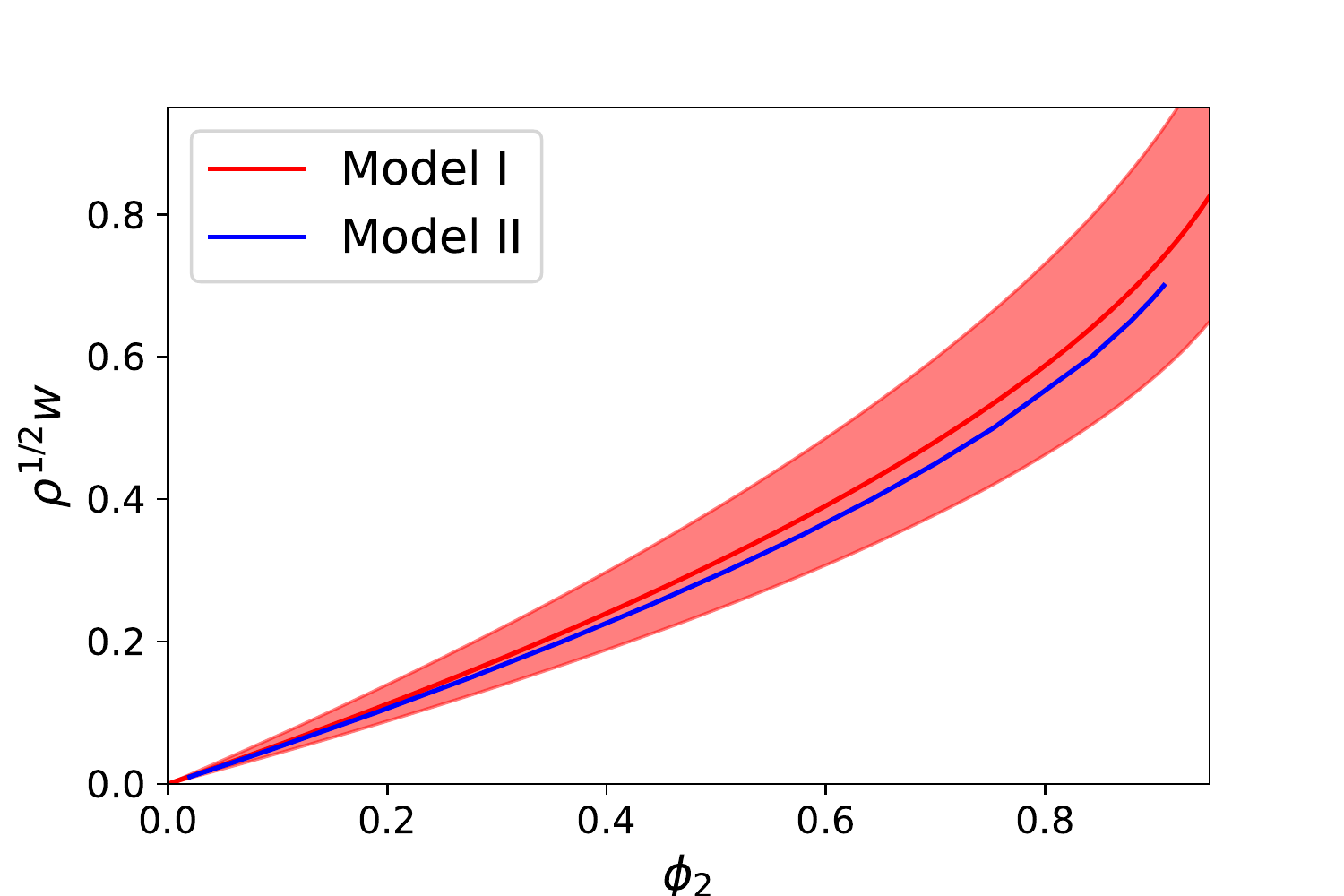}}
    \caption{
        Crack-templated networks for transparent conducting films. (a) Schematic of fabrication steps.
        (b) Two Voronoi tessellation based geometric models. Model~I is generated by shrinking each aperture so it covers a fraction of its Voronoi cell equal to the aperture area fraction of the surface, $1-\phi_2$.
        Model~II is obtained by placing wires of fixed width $w$ along each Voronoi edge, where $w$ is chosen to create an overall surface aperture area fraction of $1-\phi_2$.\cite{zeng_numerical_2020}.
        (c) Normalized wire width $\rho^{1/2} w$ versus wire area fraction $\phi_2$ for Models I and II, where $\rho$ is the number density of holes in networks.
        The red shade characterizes wire widths from Model~I (25th to 75th percentile). Model~II has a single wire width.
        \label{fig:schem}
    }
    \end{figure*} 

Myriad fabrication methods have been developed for disordered metallic networks,\cite{korte_rapid_2008,hecht2011emerging,leem2011efficient,yang2011solution,song2013highly,han_uniform_2014,ye_metal_2014,gao_physics_2016, gao_nature-inspired_2018} each producing unique geometric characteristics. For present purposes, it is helpful to distinguish between two network types. The first comprises disordered arrangements of nanowires or nanotubes, synthesized in solution and coated onto transparent substrates.\cite{korte_rapid_2008,hecht2011emerging,leem2011efficient,yang2011solution,song2013highly,ye_metal_2014,gao_physics_2016} 
Structure-property relations of these networks can be rationalized based on geometric considerations,\cite{mutiso_integrating_2013, ye_metal_2014, ocallaghan_effective_2016, he_conductivity_2018, forro_predictive_2018} though their electrical conductivity is intrinsically limited due to inter-wire contact resistance.\cite{kumar_evaluating_2016, gao_nature-inspired_2018} This limitation is addressed by the second type, so-called crack-templated networks. These networks are typically fabricated by preparing a sacrificial film that self-cracks under thermal, chemical or mechanical stress, followed by deposition of a metal film to form a seamless conductive network before cracked layer removal (Fig.~\ref{fig:schem}a).\cite{han_uniform_2014,xian_practical_2017,gao_nature-inspired_2018,jung2020recent} Despite the promise of crack-templated networks for achieving outstanding optical and electrical properties,\cite{han_uniform_2014,xian_practical_2017,zeng_numerical_2020} there has so far been limited effort directed toward understanding and predicting their performance based on geometric considerations.\cite{kumar_evaluating_2016, azani_transparent_2019, akhunzhanov_circles_2020, zeng_numerical_2020}

Here, we introduce and study two simple geometric models for crack-templated networks, characterized by disconnected polygonal holes separated by fully connected metallic wire networks. These topological properties are constructed in the models from Voronoi tessellations of disordered point configurations. The first model assigns wires of varying width to the edges of the Voronoi polygons, qualitatively mimicking a crack-templated network created by contraction of a sacrificial layer. The second model instead assigns wires of equal width to each Voronoi polygon edge, similar to a recently introduced coupled electrothermal model for transparent conducting films.\cite{zeng_numerical_2020} We compare simulated properties of these two models, including the sheet resistance and optical transmittance, to those of metallic meshes with periodic arrays of circular holes. We also discuss how the model systems perform compared to experimental results reported in the literature and describe implications of their predicted properties for applications.

To characterize the in-plane electric properties of these networks, we estimate the sheet resistance using first-passage-time simulation, an accelerated Brownian motion analysis.\cite{kim_determination_1990-1}
Different from previous approaches, including circuit analysis \cite{kumar_evaluating_2016} and the finite-element method,\cite{zeng_numerical_2020} the first-passage-time simulation enables accurate computation of the network sheet resistance, at any wire area fraction, without systematic errors due to finite resolution.\cite{kim_determination_1990-1}
Our results demonstrate that sheet resistance in these networks is nearly optimal, i.e., close to the upper Hashin-Shtrikman bound\cite{hashin_variational_1962} for two-phase composites. This observation is consistent with, but also generalizes and significantly extends, recently established results for dilute cellular networks.\cite{torquato_multifunctional_2018}

To probe the optical transmittance of these models, we employ finite-difference time-domain (FDTD) computations. The simulation results show that for both periodic and disordered networks, the normal transmittance can be enhanced beyond the classical shading limit (i.e., the area fraction of the void phase, $1-\phi_2$) for a wide range of wavelengths. However, the transmittance spectra of the crack-templated networks lack the sharp features that appear in spectra of the periodic wire networks relating to the uniform size of the latter's circular apertures and their regular spatial arrangement. The transmittance spectra of the disordered networks also lack sensitivity to incident angle, suggesting that, unlike their ordered network counterparts, they are non-iridescent, consistent with the previous observations.\cite{wen_bright_2021, chen_non-iridescent_2021}


{\it{Models.}} To understand and contextualize the effects of disorder on the properties of crack-templated networks, we consider three wire network models that differ from one another due to the distinct shapes and spatial organization of their apertures. The networks have identical film thickness and constant cross-section as a function of depth. Apertures in the reference model are circular and ordered in a periodic hexagonal array, which is commonly investigated in experiments.\cite{akinoglu_evidence_2013, qiu_metallic_2016,qiu_shape_2017} The two crack-templated models (Figs.~\ref{fig:schem}b) are deriven from Voronoi tessellations of two-dimensional point configurations. Here, the configurations represent coordinates of 100 hard-disk particle centers in a periodically-replicated simulation cell, selected from an equilibrium ensemble with particle area fraction of $0.20$. Alternative point patterns can be adopted in this framework to model a diverse range of crack microstructures.

 For the crack-templated structure we refer to as Model~I, the wire network is obtained by uniformly shrinking each polygonal template (i.e., aperture) away from the boundaries of its Voronoi cell without changing its shape, so that the resulting wire area fraction inside each cell equals the total wire area fraction of the surface, $\phi_2$. This process qualitatively mimics a crack templating process driven by a (e.g., cooling- or drying-induced) contraction of the sacrificial layer. It can be proven that the resulting geometry in Model~I is hyperuniform, indicating that---despite the disordered, isotropic structure---its long-wavelength area-fraction fluctuations are suppressed similar to ordered arrays.\cite{kim_methodology_2019}
 In this way, disordered hyperuniform patterns are intermediate between perfectly random and periodically ordered structures.\cite{Torquato2018_review}  For the network structure we term Model~II,\cite{zeng_numerical_2020} wires of identical width~$w$ are simply centered on the edges of each Voronoi cell surrounding each aperture, so that the total wire area fraction is also $\phi_2$.

The process by which Model~I networks are formed gives rise to a broad distribution of wire widths (red shaded region in Fig.~\ref{fig:schem}c).  At a given $\phi_2$, wires of Model~I are, on average, wider than those of Model~II. This is because, unlike in Model~II, the network construction for Model I preserves the shortest wires, which are thicker on average. 

We compare the electrical and optical properties of the reference model and the two crack-templated models as function of $\phi_2$ for conditions where they have identical aperture number density $\rho=\sqrt{3} \mu$m$^{-2}$ and film thickness ($80$~nm).

\begin{figure*}
\subfloat[]{\includegraphics[width=0.4\textwidth]{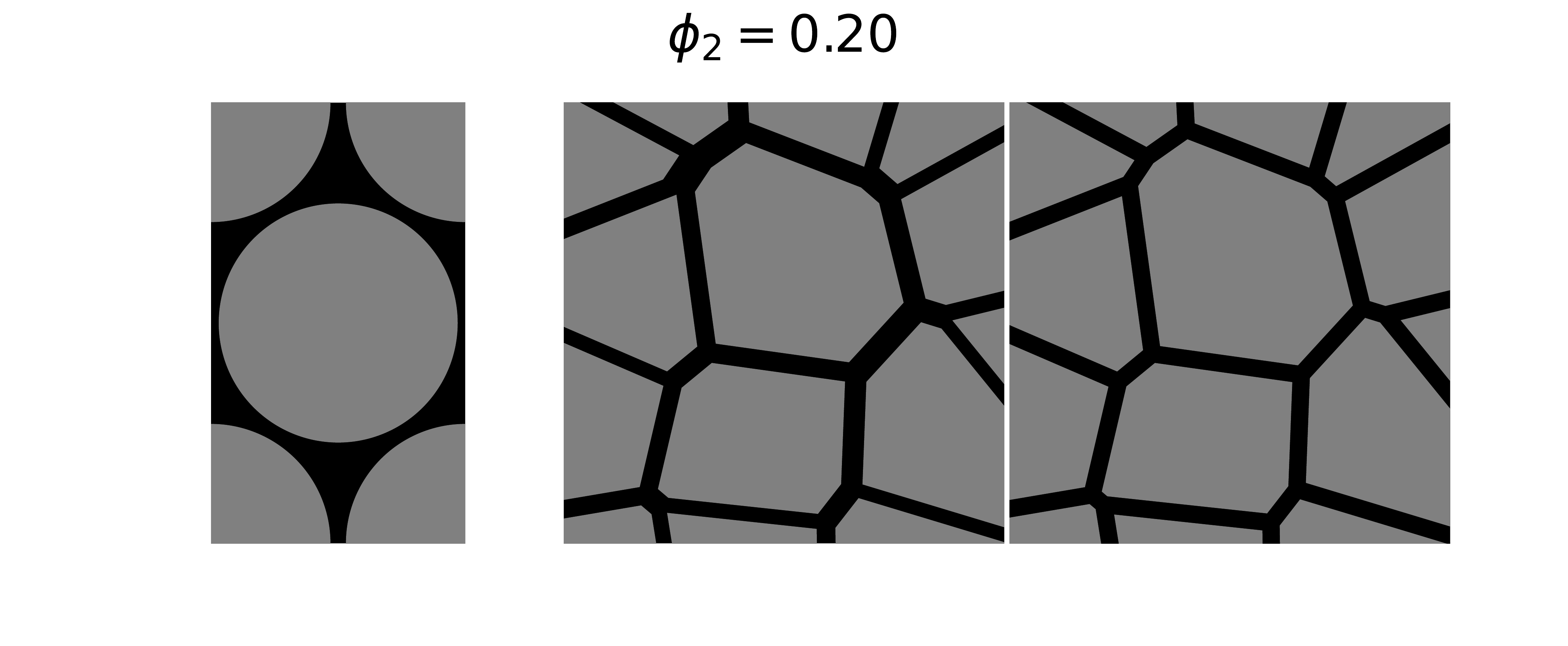}}
\vspace{-5pt}
\subfloat[]{\includegraphics[width=0.4\textwidth]{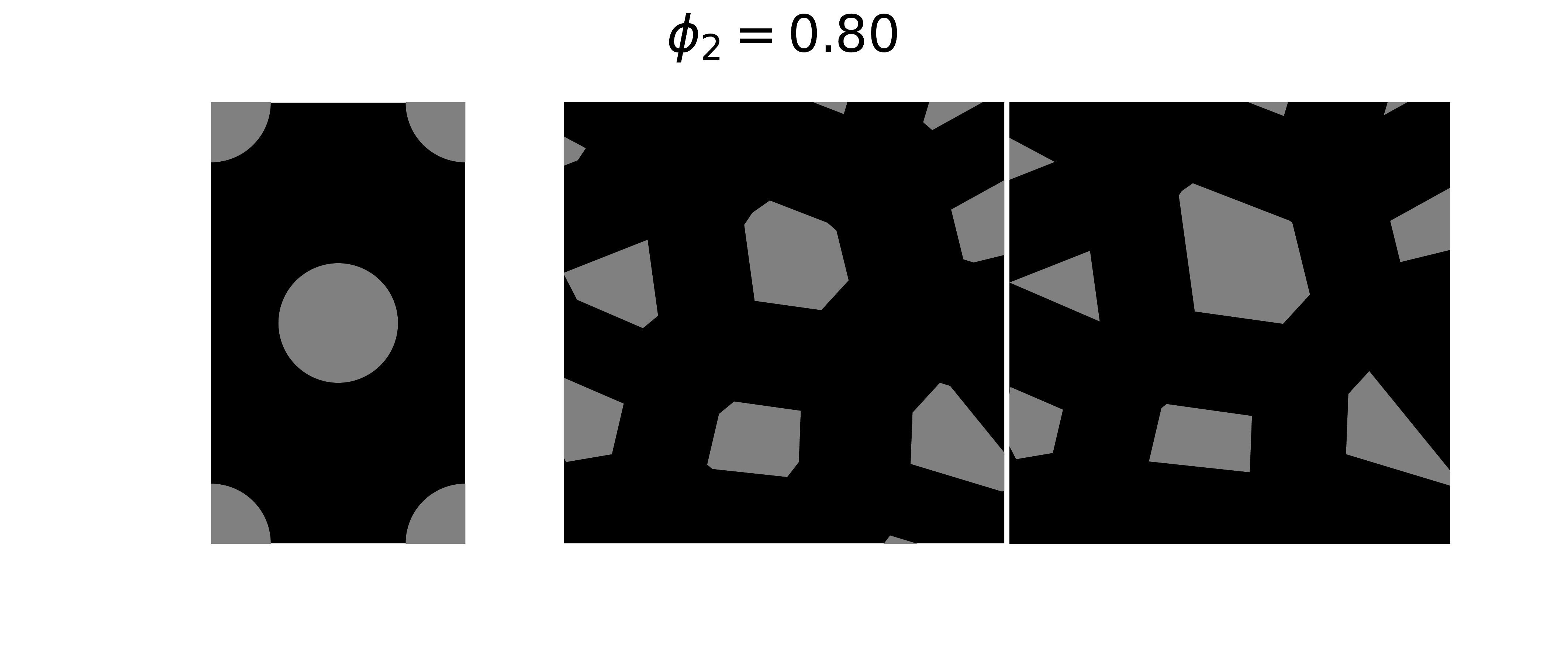}}
\vspace{-5pt}

\subfloat[]{\includegraphics[width=0.4\textwidth]{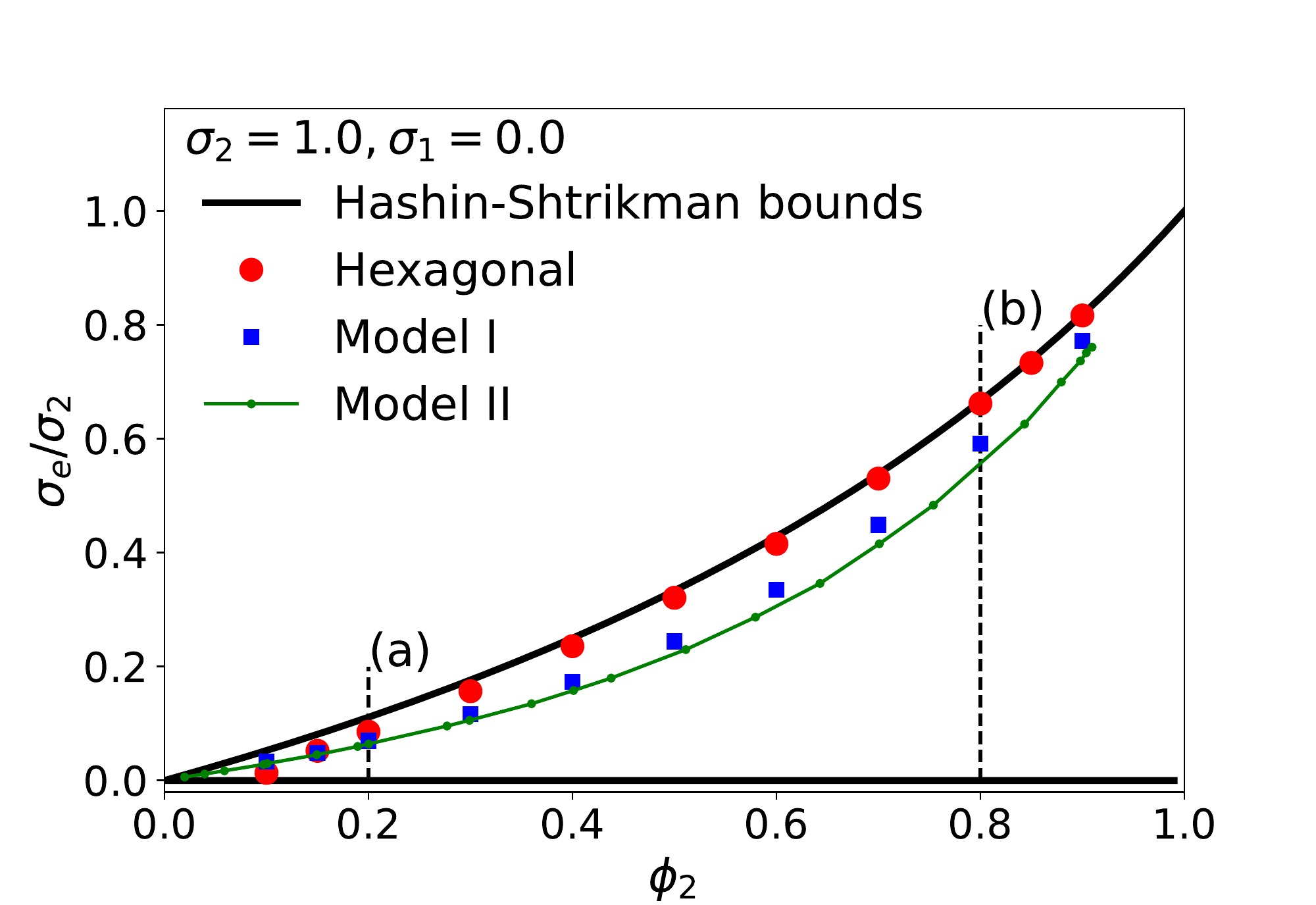}}
\subfloat[]{\includegraphics[width=0.4\textwidth]{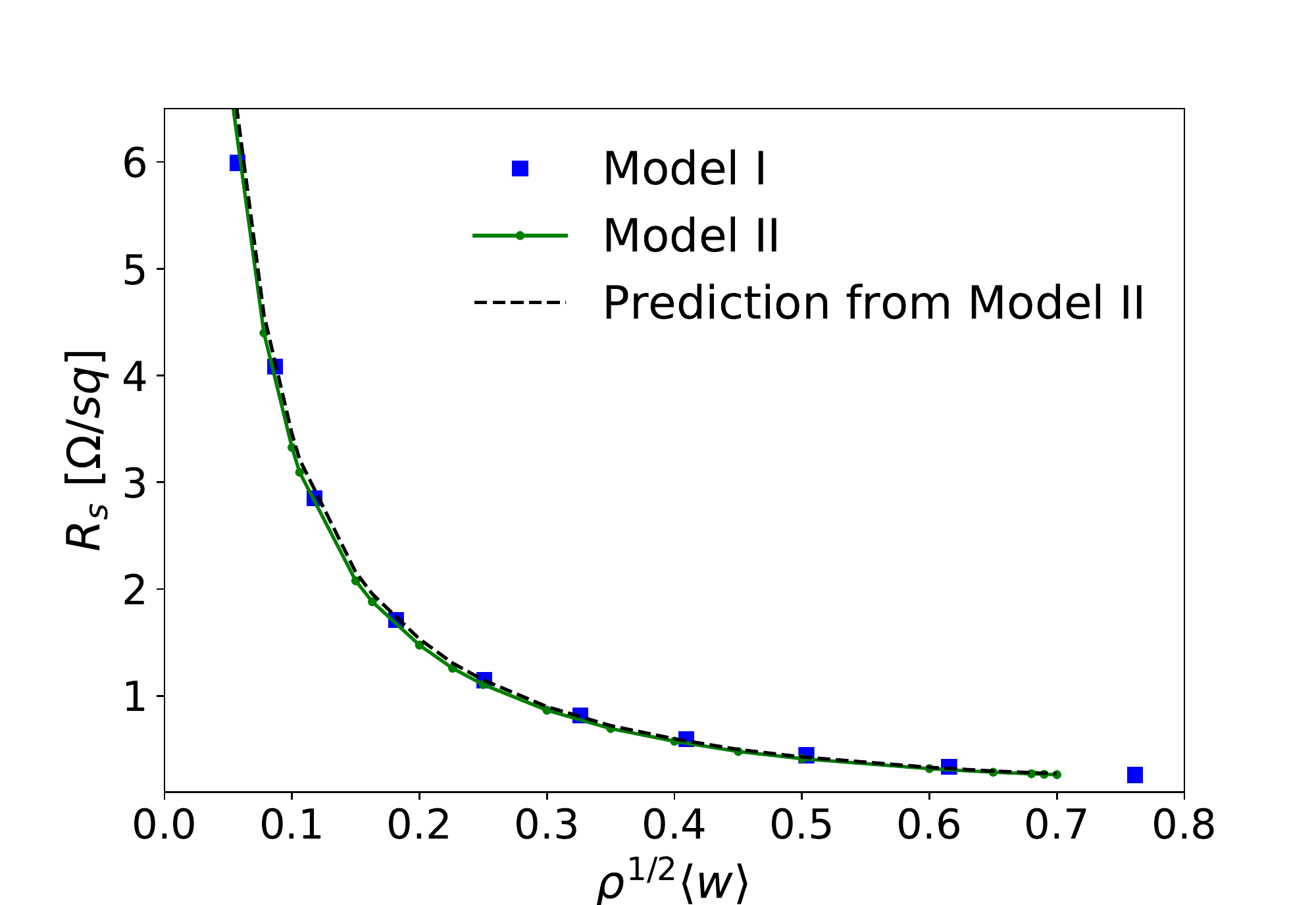}}
\caption{
    Images showing representative portions of crack-templated networks (leftmost, Model 1 and Model 2) and hexagonal array of apertures (rightmost) at (a) $\phi_2=0.2$ and (b) $\phi_2=0.8$. 
    Phases 1 and 2 (gray and black) depict the aperture (void) and metallic (wire) regions, respectively. 
    (c) Simulated effective electric conductivity $\sigma_e$ of the three network models as a function of wire network area fraction $\phi_2$ for aperture number density $\rho=\sqrt{3} \mu$m$^{-2}$ and film thickness of $80$~nm.
    The black solid line represents the upper Hashin-Shtrikman bound on $\sigma_e$, given by Eq. \eqref{eq:sigma_HS} \cite{hashin_variational_1962}.
    The two vertical dashed lines represent the area fractions depicted by images in (a) and (b).
    (d) Sheet resistance $R_s$ as a function of the normalized mean wire width $\rho^{1/2} \E{w}$ for Models I and II. 
    The black dashed curve is computed from the equality in Eq. \eqref{eq:pred-width-distribution} and the data for Model~I.
	\label{fig:electric-properties}
}
\end{figure*}

{\it{DC Electric Conductivity.}} To characterize the electric properties of the three network models, we compute the sheet resistance $R_s$ as a function of $\phi_2$ using the relation $R_s = (\sigma_e t)^{-1}$, where $\sigma_e$ is the in-plane effective conductivity, and $t$ is film thickness.
We calculate $\sigma_e$ by first-passage-time Brownian motion simulation, following the trajectories of $10^5$ non-interacting test particles in the wire phase,\cite{kim_determination_1990-1} assuming conductivities $\sigma_1 (=0)$ and $\sigma_2 (>0)$ for the aperture and wire phases (Fig.~\ref{fig:electric-properties}a and b), respectively.
This simulation technique is computationally efficient and accurate for our network models because it solely requires the boundaries of the metallic wires, which can be evaluated exactly for the polygons at each $\phi_2$.

The Hashin-Shtrikman upper bound for effective DC conductivity $\sigma_{HS}$ is given by\cite{hashin_variational_1962, torquato_random_2002}
\begin{equation}    \label{eq:sigma_HS}
    \sigma_{HS} = \E{\sigma} - \frac{\phi_1\phi_2 (\sigma_2 - \sigma_1)^2}{\E{\tilde{\sigma}}+\sigma_2 } ~~~~, 
\end{equation}
where $\E{\sigma}\equiv \phi_1\sigma_1 + \phi_2\sigma_2$ and $\E{\tilde{\sigma}}\equiv \phi_1\sigma_2 + \phi_2 \sigma_1$.
For macroscopically isotropic two-phase composites at prescribed values of conductivities and area fractions, $\sigma_{HS}$ represents the highest possible value of $\sigma_e$.\cite{hashin_variational_1962, torquato_random_2002}
We establish the periodic network's conductivity is nearly optimal over a large range of $\phi_2$,\cite{perrins_transport_1979} conditions where both model disordered networks also exhibit high conductivity, but slightly less optimal (Fig.~\ref{fig:electric-properties}c).
The near optimal conductivity of these networks is consistent with their topological properties: disconnected apertures surrounded by percolating conduction paths, without dead-ends. These trends qualitatively change for the lowest $\phi_2$ studied ($=0.10)$, where disordered films become most conductive and the ordered array approaches the geometric limit where full wire connectivity is lost.

Comparing the two crack-templated networks, Model~I is slightly more conductive than Model~II at any given $\phi_2$ (Fig.~\ref{fig:electric-properties}c). Those differences can be understood by comparing their behavior as a function of average wire width $\E{w}$ (Fig.~\ref{fig:electric-properties}d), which collapses the data to a single curve. This observation is consistent with a recent theoretical prediction comparing the sheet resistance of a network with a single wire width $R_s$ to one with a wire-width distribution $R_s^\mathrm{wd}$:\cite{kumar_evaluating_2016}
\begin{equation}    \label{eq:pred-width-distribution}
    R_s ^\mathrm{wd} = \frac{w_{AM}}{w_{GM}} R_s  \geq R_s, 
\end{equation}
where $w_{AM}$ and $w_{GM}$ are the arithmetic and geometric means of the wire widths, respectively. For Model~I and II, the data show that equality approximately holds across the entire range of $\phi_2$. 
Simulation details are provided in the Supplementary Material.

\begin{figure*}
    \subfloat[]{\includegraphics[height=0.45\textwidth]{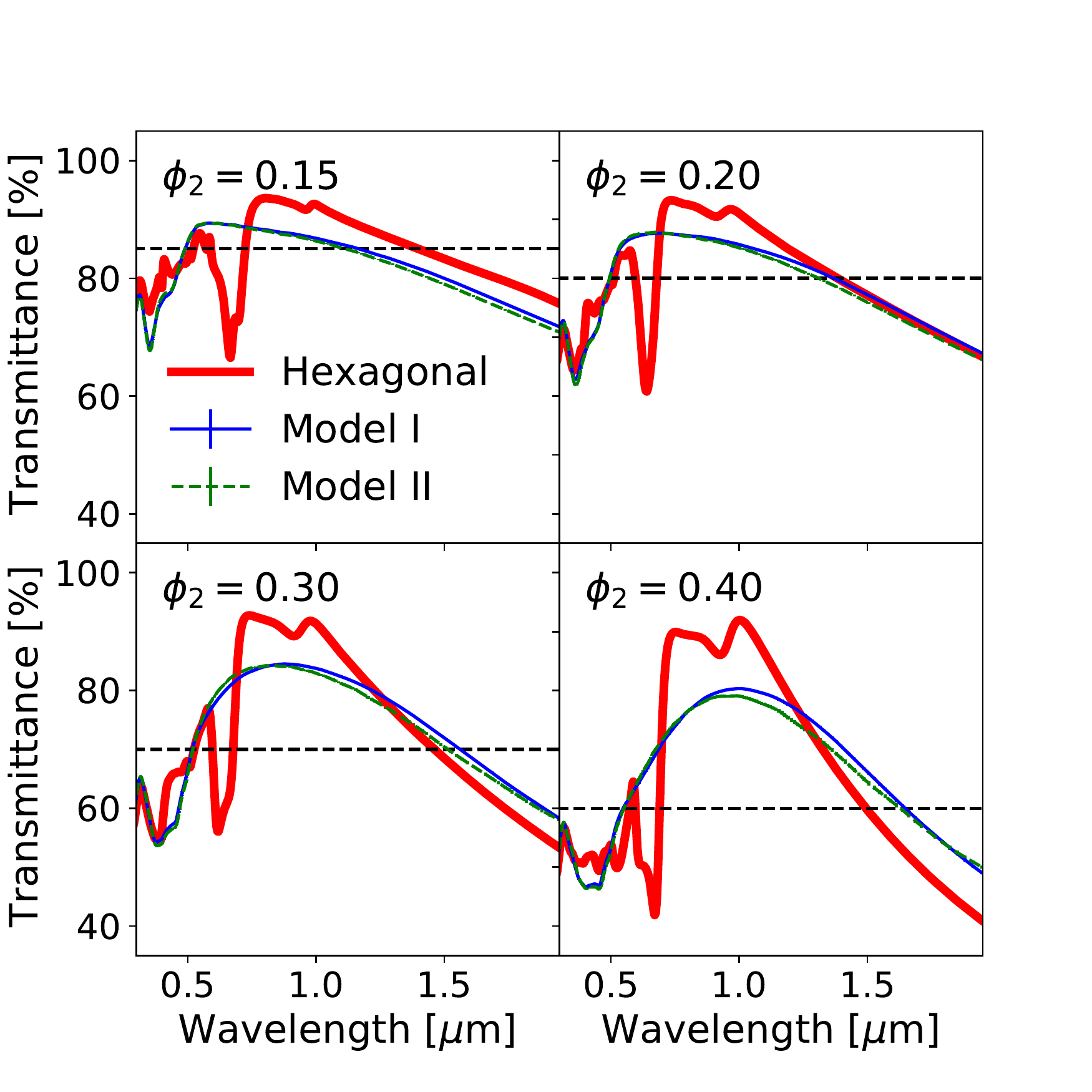}}
    \subfloat[]{\includegraphics[height=0.45\textwidth]{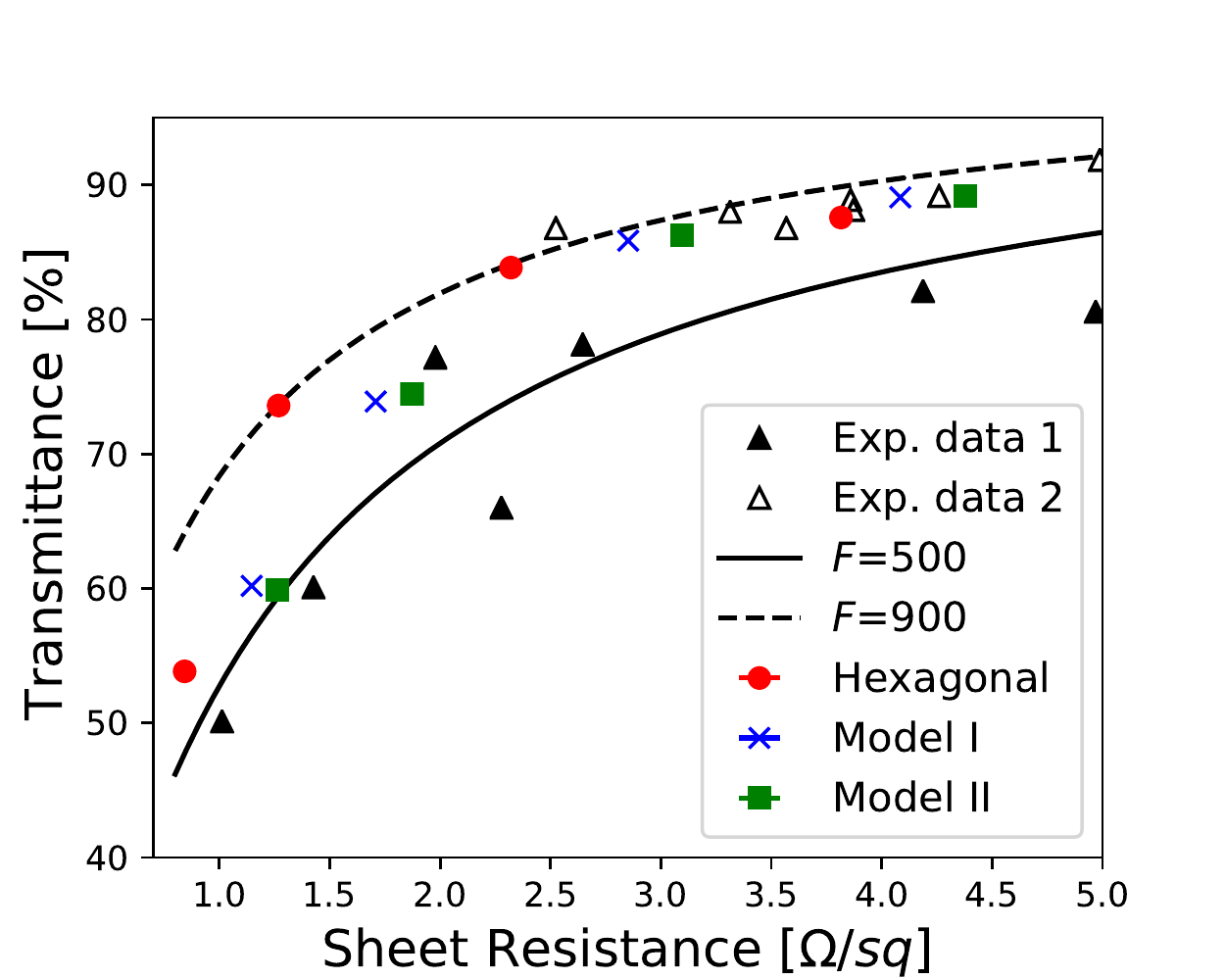}}
    \caption{
        (a) Simulated transmittance spectra at normal incidence for the three models (hexagonal array of circular holes, and two disordered, crack-templated networks) at four different values of area fraction of the silver wire phase $\phi_2$. Calculations are for aperture number density $\rho=\sqrt{3} \mu$m$^{-2}$ and film thickness of $80$~nm.
        Black dashed lines indicate the area fraction of the void phase, i.e., $1-\phi_2$, above which enhanced optical transmittance occurs.
        (b) Transmittance $T$ at 550~nm and normal incidence versus sheet resistance $R_s$ for the three simulated models and experimental data.\cite{han_optimization_2016,han_uniform_2014} Simulated models show figures of merit $500 <F < 900$ (black curves, Eq.~\ref{eq:FoM}).   
    }
    \label{fig:optical-properties}
\end{figure*}

{\it{Optical Transmittance.}} To probe the effects of aperture shape and spatial organization on optical properties,
we computed the transmittance of the ordered and crack-templated network model films via \textsc{Meep},\cite{oskooi_meep_2010} an open-source finite-difference time-domain (FDTD) simulation package,\cite{taflove_advances_2013} assuming silver as the conductive wire material (Fig.~\ref{fig:optical-properties}a). Details of the FDTD simulations are provided in the Supplementary Information.
We first consider the case of normally incident light at $\phi_2=0.15, 0.2,0.3$, and $0.4$. As expected, transmittance through the periodic networks exhibits sharp wavelength-dependent features (e.g., a sharp characteristic dip between 0.5 and 0.7 $\mu$m associated with Wood's anomaly of diffraction)\cite{ghaemi1998surface} and the enhanced optical transmission effect (i.e., transmittance $> 1-\phi_2$) between 0.8 and 1.2 $\mu$m.
Enhanced transmission, due to hole-mediating coupling between surface film plasmons and the incident light,\cite{ebbesen1998extraordinary,park2008optical} is commonly observed in metallic films perforated with hexagonal arrays of subwavelength holes.\cite{ebbesen1998extraordinary,wu2014broadband,liapis2016plasmonic,qiu_metallic_2016}
Transmittance spectra associated with the crack-templated networks lacked the pronounced dip seen in the hexagonal array meshes and instead displayed broad enhanced transmission from 0.5 $\mu$m to between 1.2 and 1.6 $\mu$m, depending on $\phi_2$. These results are for networks with mean aperture sizes of $\lesssim 1 \mu$m, and thus one can expect enhanced, wavelength-insensitive transmittance over a much broader wavelength range for crack-templated networks with apertures of larger characteristic size (e.g., from $1 - 10 \mu$m). The predicted transmittance spectra of the disordered model films are qualitatively consistent with those observed for aperiodic metallic nanomeshes,\cite{sun2014broadband} which are attractive as potential components of high efficiency photovoltaics.

For context, we also characterize all three models using a commonly employed figure of merit for transparent conductive films:\cite{gao_physics_2016, xian_practical_2017}
\begin{equation}    \label{eq:FoM}
    F = \frac{188.5}{R_s \qty(1/\sqrt{T}-1)}, 
\end{equation}
where $T$ is the transmittance evaluated at $550$~nm (Figure \ref{fig:optical-properties}b).
The performance metrics of the models track those observed experimentally in silver crack-templated films,\cite{han_optimization_2016,han_uniform_2014} and all simulated data for this study fall in the range $500 <F <900$, consistent with data from literature.\cite{gao_physics_2016}\cite{gao_physics_2016} 
The periodically ordered microstructures outperform the disordered model films at conditions of low transmittance by this metric, but this trend reverses for high transmittance. The superior performance of the crack-templated models for the latter conditions reflect their approximately optimal dc conductivity at low $\phi_2$ together with their ability to eliminate the sharp transmittance suppression that is present in the ordered films between 0.5 and 0.7 $\mu$m.

\begin{figure}
    {\includegraphics[width=0.4\textwidth]{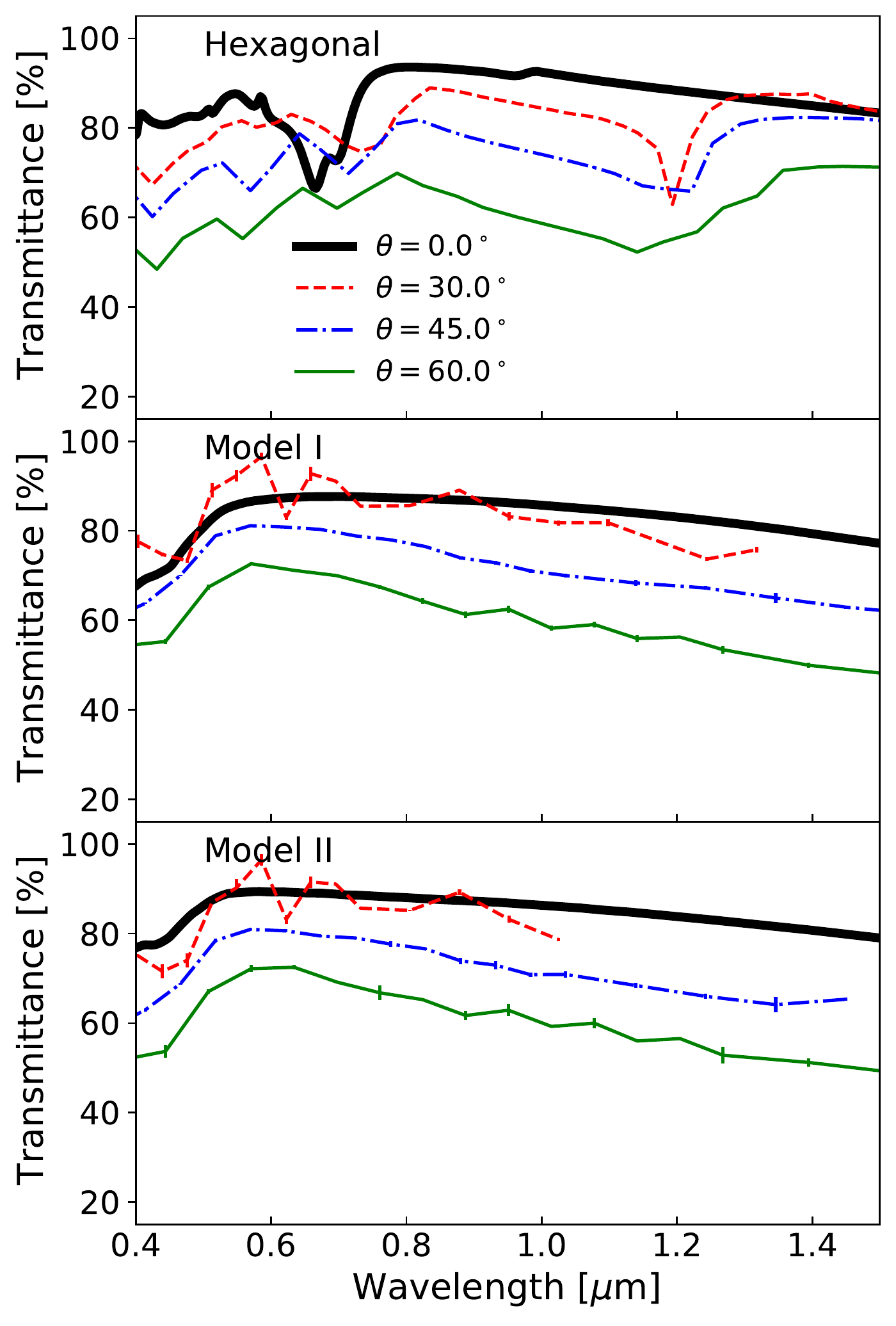}}
    \caption{
        Simulated transmittance spectra for the three models (hexagonal array of circular holes, and two disordered, crack-templated networks) at four different incident angles $\theta=0, 30, 45, 60^\circ$. Calculations are for aperture number density $\rho=\sqrt{3} \mu$m$^{-2}$ and film thickness of $80$~nm. We consider the s polarization, where the electric field of the incident light is parallel to the film surface.
        \label{fig:optical-properties2}
    }
\end{figure}

Finally, we use FDTD simulations to investigate how optical properties change with incident angle $\theta$ for the three models (Fig.~\ref{fig:optical-properties2}). For films with hexagonal arrays of perforations, the characteristic features of the transmittance spectra at normal incidence (e.g., minima and maxima) are modified considerably, and qualitatively, as a function of incident angle. By contrast, the spectral properties of the disordered films are largely angle-independent. These simulated differences are consistent with experimental comparisons between films with ordered versus quasi-2D disordered aperture structures.\cite{qiu_metallic_2016, wen_bright_2021,chen_non-iridescent_2021} The angle-dependent transmittance for ordered arrays of holes can result in films displaying bright, iridescent structural colors. This property can be challenging to control in assembly-based fabrication processes that inevitably produce defects or polycrystalline hole patterns; iridescence can also be undesirable for some applications, e.g., wide-angle viewing display-related technologies.~\cite{qiu_metallic_2016} Films with irregular aperture structures and disordered spatial patterns, like the crack-templated models explored here, may be attractive to further investigate and develop for these purposes.

{\it{Conclusions and Outlook.}} The geometric models introduced and analyzed here for transparent conducting films with
crack-templated network structures provide a simple means for understanding the interrelationships between their key structural, electric and optical properties. They establish that such network structures have near optimal DC conductivity, even surpassing that of films with hexagonal arrays of holes when the wire surface coverage is very low, i.e., in the high transmittance, high resistivity limit. This trend, together with the elimination of the sharp dip in transmittance due to Wood's anomaly, allows the cracked template model films to exhibit similar or higher transparent conductor figures-of-merit compared to films with hexagonal arrays of apertures. These models further help to quantify and explain differences in resistivity between disordered network structures and how they relate to their wire widths. Finally, the model highlights how disorder in the apertures can eliminate qualitative, incident-angle dependencies in transmittance that are characteristic of films with periodically ordered holes, an effect which has ramifications for wide-angle display applications. None of these conclusions about the properties of disordered networks appear particularly sensitive to long-wavelength fluctuations in aperture area fraction (or wire area fraction), a characteristic that distinguishes Model I and Model II. 

To probe a more diverse range of network structures, it would be straightforward to generalize the crack-templated models introduced here to include Voronoi constructions generated from other point patterns. In future work, it may be interesting to use our framework together with inverse methods\cite{yu2021engineered,kadulkar2022machine} to characterize how the wide range of possible disordered network structures impacts structure-property relations for aperiodic transparent conductors. Such studies may uncover new target crack patterns for realizing networks with novel properties or motivate the discovery of distinctive processes for fabricating them.

	We thank I.C. Kim, S. Torquato, Y. Zheng, and E. Akinoglu for helpful discussions and codes for the first-passage-time calculations and FDTD simulations, as well as D.J. Milliron for insightful feedback on the results.
	We acknowledge financial support from the Welch Foundation (Grant No. F-1696) and the Texas Advanced Computing Center (TACC) at The University of Texas at Austin for providing HPC resources.

\section*{AUTHOR DECLARATIONS\\Conflict of Interest}
The authors have no conflicts to disclose.

\section*{DATA AVAILABILITY}
The data that support the findings of this study are available from the corresponding author upon reasonable request.



\begin{thebibliography}{46}%
    \makeatletter
    \providecommand \@ifxundefined [1]{%
     \@ifx{#1\undefined}
    }%
    \providecommand \@ifnum [1]{%
     \ifnum #1\expandafter \@firstoftwo
     \else \expandafter \@secondoftwo
     \fi
    }%
    \providecommand \@ifx [1]{%
     \ifx #1\expandafter \@firstoftwo
     \else \expandafter \@secondoftwo
     \fi
    }%
    \providecommand \natexlab [1]{#1}%
    \providecommand \enquote  [1]{``#1''}%
    \providecommand \bibnamefont  [1]{#1}%
    \providecommand \bibfnamefont [1]{#1}%
    \providecommand \citenamefont [1]{#1}%
    \providecommand \href@noop [0]{\@secondoftwo}%
    \providecommand \href [0]{\begingroup \@sanitize@url \@href}%
    \providecommand \@href[1]{\@@startlink{#1}\@@href}%
    \providecommand \@@href[1]{\endgroup#1\@@endlink}%
    \providecommand \@sanitize@url [0]{\catcode `\\12\catcode `\$12\catcode
      `\&12\catcode `\#12\catcode `\^12\catcode `\_12\catcode `\%12\relax}%
    \providecommand \@@startlink[1]{}%
    \providecommand \@@endlink[0]{}%
    \providecommand \url  [0]{\begingroup\@sanitize@url \@url }%
    \providecommand \@url [1]{\endgroup\@href {#1}{\urlprefix }}%
    \providecommand \urlprefix  [0]{URL }%
    \providecommand \Eprint [0]{\href }%
    \providecommand \doibase [0]{http://dx.doi.org/}%
    \providecommand \selectlanguage [0]{\@gobble}%
    \providecommand \bibinfo  [0]{\@secondoftwo}%
    \providecommand \bibfield  [0]{\@secondoftwo}%
    \providecommand \translation [1]{[#1]}%
    \providecommand \BibitemOpen [0]{}%
    \providecommand \bibitemStop [0]{}%
    \providecommand \bibitemNoStop [0]{.\EOS\space}%
    \providecommand \EOS [0]{\spacefactor3000\relax}%
    \providecommand \BibitemShut  [1]{\csname bibitem#1\endcsname}%
    \let\auto@bib@innerbib\@empty
    \bibitem [{\citenamefont {Castell{\'o}n}\ and\ \citenamefont
      {Levy}(2018)}]{castellon2018transparent}%
      \BibitemOpen
      \bibfield  {author} {\bibinfo {author} {\bibfnamefont {E.}~\bibnamefont
      {Castell{\'o}n}}\ and\ \bibinfo {author} {\bibfnamefont {D.}~\bibnamefont
      {Levy}},\ }\href@noop {} {\emph {\bibinfo {title} {Transparent Conductive
      Materials: Materials, Synthesis, Characterization, Applications}}}\ (\bibinfo
       {publisher} {John Wiley \& Sons},\ \bibinfo {year} {2018})\BibitemShut
      {NoStop}%
    \bibitem [{\citenamefont {Han}\ \emph {et~al.}(2014)\citenamefont {Han},
      \citenamefont {Pei}, \citenamefont {Huang}, \citenamefont {Zhang},
      \citenamefont {Rong}, \citenamefont {Lin}, \citenamefont {Guo}, \citenamefont
      {Sun}, \citenamefont {Guo}, \citenamefont {Carnahan}, \citenamefont
      {Giersig}, \citenamefont {Wang}, \citenamefont {Gao}, \citenamefont {Ren},\
      and\ \citenamefont {Kempa}}]{han_uniform_2014}%
      \BibitemOpen
      \bibfield  {author} {\bibinfo {author} {\bibfnamefont {B.}~\bibnamefont
      {Han}}, \bibinfo {author} {\bibfnamefont {K.}~\bibnamefont {Pei}}, \bibinfo
      {author} {\bibfnamefont {Y.}~\bibnamefont {Huang}}, \bibinfo {author}
      {\bibfnamefont {X.}~\bibnamefont {Zhang}}, \bibinfo {author} {\bibfnamefont
      {Q.}~\bibnamefont {Rong}}, \bibinfo {author} {\bibfnamefont {Q.}~\bibnamefont
      {Lin}}, \bibinfo {author} {\bibfnamefont {Y.}~\bibnamefont {Guo}}, \bibinfo
      {author} {\bibfnamefont {T.}~\bibnamefont {Sun}}, \bibinfo {author}
      {\bibfnamefont {C.}~\bibnamefont {Guo}}, \bibinfo {author} {\bibfnamefont
      {D.}~\bibnamefont {Carnahan}}, \bibinfo {author} {\bibfnamefont
      {M.}~\bibnamefont {Giersig}}, \bibinfo {author} {\bibfnamefont
      {Y.}~\bibnamefont {Wang}}, \bibinfo {author} {\bibfnamefont {J.}~\bibnamefont
      {Gao}}, \bibinfo {author} {\bibfnamefont {Z.}~\bibnamefont {Ren}}, \ and\
      \bibinfo {author} {\bibfnamefont {K.}~\bibnamefont {Kempa}},\ }\bibfield
      {title} {\enquote {\bibinfo {title} {Uniform {{Self}}-{{Forming Metallic
      Network}} as a {{High}}-{{Performance Transparent Conductive Electrode}}},}\
      }\href {\doibase 10.1002/adma.201302950} {\bibfield  {journal} {\bibinfo
      {journal} {Adv. Mater.}\ }\textbf {\bibinfo {volume} {26}},\ \bibinfo {pages}
      {873--877} (\bibinfo {year} {2014})}\BibitemShut {NoStop}%
    \bibitem [{\citenamefont {Ye}\ \emph {et~al.}(2014)\citenamefont {Ye},
      \citenamefont {Rathmell}, \citenamefont {Chen}, \citenamefont {Stewart},\
      and\ \citenamefont {Wiley}}]{ye_metal_2014}%
      \BibitemOpen
      \bibfield  {author} {\bibinfo {author} {\bibfnamefont {S.}~\bibnamefont
      {Ye}}, \bibinfo {author} {\bibfnamefont {A.~R.}\ \bibnamefont {Rathmell}},
      \bibinfo {author} {\bibfnamefont {Z.}~\bibnamefont {Chen}}, \bibinfo {author}
      {\bibfnamefont {I.~E.}\ \bibnamefont {Stewart}}, \ and\ \bibinfo {author}
      {\bibfnamefont {B.~J.}\ \bibnamefont {Wiley}},\ }\bibfield  {title} {\enquote
      {\bibinfo {title} {Metal {{Nanowire Networks}}: The {{Next Generation}} of
      {{Transparent Conductors}}},}\ }\href {\doibase 10.1002/adma.201402710}
      {\bibfield  {journal} {\bibinfo  {journal} {Adv. Mater.}\ }\textbf {\bibinfo
      {volume} {26}},\ \bibinfo {pages} {6670--6687} (\bibinfo {year}
      {2014})}\BibitemShut {NoStop}%
    \bibitem [{\citenamefont {Gao}\ \emph {et~al.}(2016)\citenamefont {Gao},
      \citenamefont {Kempa}, \citenamefont {Giersig}, \citenamefont {Akinoglu},
      \citenamefont {Han},\ and\ \citenamefont {Li}}]{gao_physics_2016}%
      \BibitemOpen
      \bibfield  {author} {\bibinfo {author} {\bibfnamefont {J.}~\bibnamefont
      {Gao}}, \bibinfo {author} {\bibfnamefont {K.}~\bibnamefont {Kempa}}, \bibinfo
      {author} {\bibfnamefont {M.}~\bibnamefont {Giersig}}, \bibinfo {author}
      {\bibfnamefont {E.~M.}\ \bibnamefont {Akinoglu}}, \bibinfo {author}
      {\bibfnamefont {B.}~\bibnamefont {Han}}, \ and\ \bibinfo {author}
      {\bibfnamefont {R.}~\bibnamefont {Li}},\ }\bibfield  {title} {\enquote
      {\bibinfo {title} {Physics of transparent conductors},}\ }\href {\doibase
      10.1080/00018732.2016.1226804} {\bibfield  {journal} {\bibinfo  {journal}
      {Adv. Phys.}\ }\textbf {\bibinfo {volume} {65}},\ \bibinfo {pages} {553--617}
      (\bibinfo {year} {2016})}\BibitemShut {NoStop}%
    \bibitem [{\citenamefont {Gao}\ \emph {et~al.}(2018)\citenamefont {Gao},
      \citenamefont {Xian}, \citenamefont {Zhou}, \citenamefont {Liu},\ and\
      \citenamefont {Kempa}}]{gao_nature-inspired_2018}%
      \BibitemOpen
      \bibfield  {author} {\bibinfo {author} {\bibfnamefont {J.}~\bibnamefont
      {Gao}}, \bibinfo {author} {\bibfnamefont {Z.}~\bibnamefont {Xian}}, \bibinfo
      {author} {\bibfnamefont {G.}~\bibnamefont {Zhou}}, \bibinfo {author}
      {\bibfnamefont {J.-M.}\ \bibnamefont {Liu}}, \ and\ \bibinfo {author}
      {\bibfnamefont {K.}~\bibnamefont {Kempa}},\ }\bibfield  {title} {\enquote
      {\bibinfo {title} {Nature-{{Inspired Metallic Networks}} for {{Transparent
      Electrodes}}},}\ }\href {\doibase 10.1002/adfm.201705023} {\bibfield
      {journal} {\bibinfo  {journal} {Adv. Funct. Mater.}\ }\textbf {\bibinfo
      {volume} {28}},\ \bibinfo {pages} {1705023} (\bibinfo {year}
      {2018})}\BibitemShut {NoStop}%
    \bibitem [{\citenamefont {Jung}\ \emph {et~al.}(2020)\citenamefont {Jung},
      \citenamefont {Kim}, \citenamefont {Suh}, \citenamefont {Hong}, \citenamefont
      {Yeo},\ and\ \citenamefont {Ko}}]{jung2020recent}%
      \BibitemOpen
      \bibfield  {author} {\bibinfo {author} {\bibfnamefont {J.}~\bibnamefont
      {Jung}}, \bibinfo {author} {\bibfnamefont {K.~K.}\ \bibnamefont {Kim}},
      \bibinfo {author} {\bibfnamefont {Y.~D.}\ \bibnamefont {Suh}}, \bibinfo
      {author} {\bibfnamefont {S.}~\bibnamefont {Hong}}, \bibinfo {author}
      {\bibfnamefont {J.}~\bibnamefont {Yeo}}, \ and\ \bibinfo {author}
      {\bibfnamefont {S.~H.}\ \bibnamefont {Ko}},\ }\bibfield  {title} {\enquote
      {\bibinfo {title} {Recent progress in controlled nano/micro cracking as an
      alternative nano-patterning method for functional applications},}\
      }\href@noop {} {\bibfield  {journal} {\bibinfo  {journal} {Nanoscale
      Horizons}\ }\textbf {\bibinfo {volume} {5}},\ \bibinfo {pages} {1036--1049}
      (\bibinfo {year} {2020})}\BibitemShut {NoStop}%
    \bibitem [{\citenamefont {Muzzillo}\ \emph {et~al.}(2020)\citenamefont
      {Muzzillo}, \citenamefont {Wong}, \citenamefont {Mansfield}, \citenamefont
      {Simon},\ and\ \citenamefont {Ptak}}]{muzzillo2020patterning}%
      \BibitemOpen
      \bibfield  {author} {\bibinfo {author} {\bibfnamefont {C.~P.}\ \bibnamefont
      {Muzzillo}}, \bibinfo {author} {\bibfnamefont {E.}~\bibnamefont {Wong}},
      \bibinfo {author} {\bibfnamefont {L.~M.}\ \bibnamefont {Mansfield}}, \bibinfo
      {author} {\bibfnamefont {J.}~\bibnamefont {Simon}}, \ and\ \bibinfo {author}
      {\bibfnamefont {A.~J.}\ \bibnamefont {Ptak}},\ }\bibfield  {title} {\enquote
      {\bibinfo {title} {Patterning metal grids for gaas solar cells with cracked
      film lithography: Quantifying the cost/performance tradeoff},}\ }\href@noop
      {} {\bibfield  {journal} {\bibinfo  {journal} {ACS Appl. Mater. Int.}\
      }\textbf {\bibinfo {volume} {12}},\ \bibinfo {pages} {41471--41476} (\bibinfo
      {year} {2020})}\BibitemShut {NoStop}%
    \bibitem [{\citenamefont {Qiu}\ \emph {et~al.}(2016)\citenamefont {Qiu},
      \citenamefont {Luo}, \citenamefont {Ali}, \citenamefont {Jaatinen},
      \citenamefont {Wang},\ and\ \citenamefont {Wang}}]{qiu_metallic_2016}%
      \BibitemOpen
      \bibfield  {author} {\bibinfo {author} {\bibfnamefont {T.}~\bibnamefont
      {Qiu}}, \bibinfo {author} {\bibfnamefont {B.}~\bibnamefont {Luo}}, \bibinfo
      {author} {\bibfnamefont {F.}~\bibnamefont {Ali}}, \bibinfo {author}
      {\bibfnamefont {E.}~\bibnamefont {Jaatinen}}, \bibinfo {author}
      {\bibfnamefont {L.}~\bibnamefont {Wang}}, \ and\ \bibinfo {author}
      {\bibfnamefont {H.}~\bibnamefont {Wang}},\ }\bibfield  {title} {\enquote
      {\bibinfo {title} {Metallic {{Nanomesh}} with {{Disordered Dual-Size
      Apertures As Wide-Viewing-Angle Transparent Conductive Electrode}}},}\ }\href
      {\doibase 10.1021/acsami.6b08173} {\bibfield  {journal} {\bibinfo  {journal}
      {ACS Appl. Mater. Int.}\ }\textbf {\bibinfo {volume} {8}},\ \bibinfo {pages}
      {22768--22773} (\bibinfo {year} {2016})}\BibitemShut {NoStop}%
    \bibitem [{\citenamefont {Wen}\ \emph {et~al.}(2021)\citenamefont {Wen},
      \citenamefont {Lu}, \citenamefont {Wei}, \citenamefont {Li}, \citenamefont
      {Li},\ and\ \citenamefont {Yang}}]{wen_bright_2021}%
      \BibitemOpen
      \bibfield  {author} {\bibinfo {author} {\bibfnamefont {X.}~\bibnamefont
      {Wen}}, \bibinfo {author} {\bibfnamefont {X.}~\bibnamefont {Lu}}, \bibinfo
      {author} {\bibfnamefont {C.}~\bibnamefont {Wei}}, \bibinfo {author}
      {\bibfnamefont {J.}~\bibnamefont {Li}}, \bibinfo {author} {\bibfnamefont
      {Y.}~\bibnamefont {Li}}, \ and\ \bibinfo {author} {\bibfnamefont
      {S.}~\bibnamefont {Yang}},\ }\bibfield  {title} {\enquote {\bibinfo {title}
      {Bright, {{Angle-Independent}}, {{Solvent-Responsive}}, and {{Structurally
      Colored Coatings}} and {{Rewritable Photonic Paper Based}} on
      {{High-Refractive-Index Colloidal Quasi-Amorphous Arrays}}},}\ }\href
      {\doibase 10.1021/acsanm.1c02283} {\bibfield  {journal} {\bibinfo  {journal}
      {ACS Appl. Nano Mater.}\ }\textbf {\bibinfo {volume} {4}},\ \bibinfo {pages}
      {9855--9865} (\bibinfo {year} {2021})}\BibitemShut {NoStop}%
    \bibitem [{\citenamefont {Chen}\ \emph {et~al.}(2021)\citenamefont {Chen},
      \citenamefont {He}, \citenamefont {Chen}, \citenamefont {Huang},
      \citenamefont {Li}, \citenamefont {Cui}, \citenamefont {Yuan},\ and\
      \citenamefont {Ge}}]{chen_non-iridescent_2021}%
      \BibitemOpen
      \bibfield  {author} {\bibinfo {author} {\bibfnamefont {X.}~\bibnamefont
      {Chen}}, \bibinfo {author} {\bibfnamefont {Y.}~\bibnamefont {He}}, \bibinfo
      {author} {\bibfnamefont {X.}~\bibnamefont {Chen}}, \bibinfo {author}
      {\bibfnamefont {C.}~\bibnamefont {Huang}}, \bibinfo {author} {\bibfnamefont
      {Y.}~\bibnamefont {Li}}, \bibinfo {author} {\bibfnamefont {Y.}~\bibnamefont
      {Cui}}, \bibinfo {author} {\bibfnamefont {C.}~\bibnamefont {Yuan}}, \ and\
      \bibinfo {author} {\bibfnamefont {H.}~\bibnamefont {Ge}},\ }\bibfield
      {title} {\enquote {\bibinfo {title} {Non-{{Iridescent Metal Nanomesh}} with
      {{Disordered Nanoapertures Fabricated}} by {{Phase Separation Lithography}}
      of {{Polymer Blend}} as {{Transparent Conductive Film}}},}\ }\href {\doibase
      10.3390/ma14040867} {\bibfield  {journal} {\bibinfo  {journal} {Materials}\
      }\textbf {\bibinfo {volume} {14}},\ \bibinfo {pages} {867} (\bibinfo {year}
      {2021})}\BibitemShut {NoStop}%
    \bibitem [{\citenamefont {Suh}\ \emph {et~al.}(2016)\citenamefont {Suh},
      \citenamefont {Hong}, \citenamefont {Lee}, \citenamefont {Lee}, \citenamefont
      {Jung}, \citenamefont {Kwon}, \citenamefont {Moon}, \citenamefont {Won},
      \citenamefont {Shin}, \citenamefont {Yeo} \emph {et~al.}}]{suh2016random}%
      \BibitemOpen
      \bibfield  {author} {\bibinfo {author} {\bibfnamefont {Y.~D.}\ \bibnamefont
      {Suh}}, \bibinfo {author} {\bibfnamefont {S.}~\bibnamefont {Hong}}, \bibinfo
      {author} {\bibfnamefont {J.}~\bibnamefont {Lee}}, \bibinfo {author}
      {\bibfnamefont {H.}~\bibnamefont {Lee}}, \bibinfo {author} {\bibfnamefont
      {S.}~\bibnamefont {Jung}}, \bibinfo {author} {\bibfnamefont {J.}~\bibnamefont
      {Kwon}}, \bibinfo {author} {\bibfnamefont {H.}~\bibnamefont {Moon}}, \bibinfo
      {author} {\bibfnamefont {P.}~\bibnamefont {Won}}, \bibinfo {author}
      {\bibfnamefont {J.}~\bibnamefont {Shin}}, \bibinfo {author} {\bibfnamefont
      {J.}~\bibnamefont {Yeo}},  \emph {et~al.},\ }\bibfield  {title} {\enquote
      {\bibinfo {title} {Random nanocrack, assisted metal nanowire-bundled network
      fabrication for a highly flexible and transparent conductor},}\ }\href@noop
      {} {\bibfield  {journal} {\bibinfo  {journal} {RSC Adv.}\ }\textbf {\bibinfo
      {volume} {6}},\ \bibinfo {pages} {57434--57440} (\bibinfo {year}
      {2016})}\BibitemShut {NoStop}%
    \bibitem [{\citenamefont {Jung}\ \emph {et~al.}(2019)\citenamefont {Jung},
      \citenamefont {Cho}, \citenamefont {Choi}, \citenamefont {Kim}, \citenamefont
      {Kwon}, \citenamefont {Shin}, \citenamefont {Hong}, \citenamefont {Kim},
      \citenamefont {Yoon}, \citenamefont {Lee} \emph {et~al.}}]{jung2019moire}%
      \BibitemOpen
      \bibfield  {author} {\bibinfo {author} {\bibfnamefont {J.}~\bibnamefont
      {Jung}}, \bibinfo {author} {\bibfnamefont {H.}~\bibnamefont {Cho}}, \bibinfo
      {author} {\bibfnamefont {S.~H.}\ \bibnamefont {Choi}}, \bibinfo {author}
      {\bibfnamefont {D.}~\bibnamefont {Kim}}, \bibinfo {author} {\bibfnamefont
      {J.}~\bibnamefont {Kwon}}, \bibinfo {author} {\bibfnamefont {J.}~\bibnamefont
      {Shin}}, \bibinfo {author} {\bibfnamefont {S.}~\bibnamefont {Hong}}, \bibinfo
      {author} {\bibfnamefont {H.}~\bibnamefont {Kim}}, \bibinfo {author}
      {\bibfnamefont {Y.}~\bibnamefont {Yoon}}, \bibinfo {author} {\bibfnamefont
      {J.}~\bibnamefont {Lee}},  \emph {et~al.},\ }\bibfield  {title} {\enquote
      {\bibinfo {title} {Moir{\'e}-free imperceptible and flexible random metal
      grid electrodes with large figure-of-merit by photonic sintering control of
      copper nanoparticles},}\ }\href@noop {} {\bibfield  {journal} {\bibinfo
      {journal} {ACS Appl. Mater. Int.}\ }\textbf {\bibinfo {volume} {11}},\
      \bibinfo {pages} {15773--15780} (\bibinfo {year} {2019})}\BibitemShut
      {NoStop}%
    \bibitem [{\citenamefont {Zeng}, \citenamefont {Wang},\ and\ \citenamefont
      {Gao}(2020)}]{zeng_numerical_2020}%
      \BibitemOpen
      \bibfield  {author} {\bibinfo {author} {\bibfnamefont {Z.}~\bibnamefont
      {Zeng}}, \bibinfo {author} {\bibfnamefont {C.}~\bibnamefont {Wang}}, \ and\
      \bibinfo {author} {\bibfnamefont {J.}~\bibnamefont {Gao}},\ }\bibfield
      {title} {\enquote {\bibinfo {title} {Numerical simulation and optimization of
      metallic network for highly efficient transparent conductive films},}\ }\href
      {\doibase 10.1063/1.5141162} {\bibfield  {journal} {\bibinfo  {journal} {J.
      Appl. Phys.}\ }\textbf {\bibinfo {volume} {127}},\ \bibinfo {pages} {065104}
      (\bibinfo {year} {2020})}\BibitemShut {NoStop}%
    \bibitem [{\citenamefont {Korte}, \citenamefont {Skrabalak},\ and\
      \citenamefont {Xia}(2008)}]{korte_rapid_2008}%
      \BibitemOpen
      \bibfield  {author} {\bibinfo {author} {\bibfnamefont {K.~E.}\ \bibnamefont
      {Korte}}, \bibinfo {author} {\bibfnamefont {S.~E.}\ \bibnamefont
      {Skrabalak}}, \ and\ \bibinfo {author} {\bibfnamefont {Y.}~\bibnamefont
      {Xia}},\ }\bibfield  {title} {\enquote {\bibinfo {title} {Rapid synthesis of
      silver nanowires through a {{CuCl-}} or {{CuCl2-mediated}} polyol process},}\
      }\href {\doibase 10.1039/B714072J} {\bibfield  {journal} {\bibinfo  {journal}
      {J. Mater. Chem.}\ }\textbf {\bibinfo {volume} {18}},\ \bibinfo {pages}
      {437--441} (\bibinfo {year} {2008})}\BibitemShut {NoStop}%
    \bibitem [{\citenamefont {Hecht}, \citenamefont {Hu},\ and\ \citenamefont
      {Irvin}(2011)}]{hecht2011emerging}%
      \BibitemOpen
      \bibfield  {author} {\bibinfo {author} {\bibfnamefont {D.~S.}\ \bibnamefont
      {Hecht}}, \bibinfo {author} {\bibfnamefont {L.}~\bibnamefont {Hu}}, \ and\
      \bibinfo {author} {\bibfnamefont {G.}~\bibnamefont {Irvin}},\ }\bibfield
      {title} {\enquote {\bibinfo {title} {Emerging transparent electrodes based on
      thin films of carbon nanotubes, graphene, and metallic nanostructures},}\
      }\href@noop {} {\bibfield  {journal} {\bibinfo  {journal} {Adv. Mater.}\
      }\textbf {\bibinfo {volume} {23}},\ \bibinfo {pages} {1482--1513} (\bibinfo
      {year} {2011})}\BibitemShut {NoStop}%
    \bibitem [{\citenamefont {Leem}\ \emph {et~al.}(2011)\citenamefont {Leem},
      \citenamefont {Edwards}, \citenamefont {Faist}, \citenamefont {Nelson},
      \citenamefont {Bradley},\ and\ \citenamefont {De~Mello}}]{leem2011efficient}%
      \BibitemOpen
      \bibfield  {author} {\bibinfo {author} {\bibfnamefont {D.-S.}\ \bibnamefont
      {Leem}}, \bibinfo {author} {\bibfnamefont {A.}~\bibnamefont {Edwards}},
      \bibinfo {author} {\bibfnamefont {M.}~\bibnamefont {Faist}}, \bibinfo
      {author} {\bibfnamefont {J.}~\bibnamefont {Nelson}}, \bibinfo {author}
      {\bibfnamefont {D.~D.}\ \bibnamefont {Bradley}}, \ and\ \bibinfo {author}
      {\bibfnamefont {J.~C.}\ \bibnamefont {De~Mello}},\ }\bibfield  {title}
      {\enquote {\bibinfo {title} {Efficient organic solar cells with
      solution-processed silver nanowire electrodes},}\ }\href@noop {} {\bibfield
      {journal} {\bibinfo  {journal} {Adv. Mater.}\ }\textbf {\bibinfo {volume}
      {23}},\ \bibinfo {pages} {4371--4375} (\bibinfo {year} {2011})}\BibitemShut
      {NoStop}%
    \bibitem [{\citenamefont {Yang}\ \emph {et~al.}(2011)\citenamefont {Yang},
      \citenamefont {Zhang}, \citenamefont {Zhou}, \citenamefont {Price},
      \citenamefont {Wiley},\ and\ \citenamefont {You}}]{yang2011solution}%
      \BibitemOpen
      \bibfield  {author} {\bibinfo {author} {\bibfnamefont {L.}~\bibnamefont
      {Yang}}, \bibinfo {author} {\bibfnamefont {T.}~\bibnamefont {Zhang}},
      \bibinfo {author} {\bibfnamefont {H.}~\bibnamefont {Zhou}}, \bibinfo {author}
      {\bibfnamefont {S.~C.}\ \bibnamefont {Price}}, \bibinfo {author}
      {\bibfnamefont {B.~J.}\ \bibnamefont {Wiley}}, \ and\ \bibinfo {author}
      {\bibfnamefont {W.}~\bibnamefont {You}},\ }\bibfield  {title} {\enquote
      {\bibinfo {title} {Solution-processed flexible polymer solar cells with
      silver nanowire electrodes},}\ }\href@noop {} {\bibfield  {journal} {\bibinfo
       {journal} {ACS Appl. Mater. Int.}\ }\textbf {\bibinfo {volume} {3}},\
      \bibinfo {pages} {4075--4084} (\bibinfo {year} {2011})}\BibitemShut {NoStop}%
    \bibitem [{\citenamefont {Song}\ \emph {et~al.}(2013)\citenamefont {Song},
      \citenamefont {You}, \citenamefont {Lim}, \citenamefont {Park}, \citenamefont
      {Jung}, \citenamefont {Kim}, \citenamefont {Kim}, \citenamefont {Kim},
      \citenamefont {Kim}, \citenamefont {Park} \emph {et~al.}}]{song2013highly}%
      \BibitemOpen
      \bibfield  {author} {\bibinfo {author} {\bibfnamefont {M.}~\bibnamefont
      {Song}}, \bibinfo {author} {\bibfnamefont {D.~S.}\ \bibnamefont {You}},
      \bibinfo {author} {\bibfnamefont {K.}~\bibnamefont {Lim}}, \bibinfo {author}
      {\bibfnamefont {S.}~\bibnamefont {Park}}, \bibinfo {author} {\bibfnamefont
      {S.}~\bibnamefont {Jung}}, \bibinfo {author} {\bibfnamefont {C.~S.}\
      \bibnamefont {Kim}}, \bibinfo {author} {\bibfnamefont {D.-H.}\ \bibnamefont
      {Kim}}, \bibinfo {author} {\bibfnamefont {D.-G.}\ \bibnamefont {Kim}},
      \bibinfo {author} {\bibfnamefont {J.-K.}\ \bibnamefont {Kim}}, \bibinfo
      {author} {\bibfnamefont {J.}~\bibnamefont {Park}},  \emph {et~al.},\
      }\bibfield  {title} {\enquote {\bibinfo {title} {Highly efficient and
      bendable organic solar cells with solution-processed silver nanowire
      electrodes},}\ }\href@noop {} {\bibfield  {journal} {\bibinfo  {journal}
      {Adv. Funct. Mater.}\ }\textbf {\bibinfo {volume} {23}},\ \bibinfo {pages}
      {4177--4184} (\bibinfo {year} {2013})}\BibitemShut {NoStop}%
    \bibitem [{\citenamefont {Mutiso}\ \emph {et~al.}(2013)\citenamefont {Mutiso},
      \citenamefont {Sherrott}, \citenamefont {Rathmell}, \citenamefont {Wiley},\
      and\ \citenamefont {Winey}}]{mutiso_integrating_2013}%
      \BibitemOpen
      \bibfield  {author} {\bibinfo {author} {\bibfnamefont {R.~M.}\ \bibnamefont
      {Mutiso}}, \bibinfo {author} {\bibfnamefont {M.~C.}\ \bibnamefont
      {Sherrott}}, \bibinfo {author} {\bibfnamefont {A.~R.}\ \bibnamefont
      {Rathmell}}, \bibinfo {author} {\bibfnamefont {B.~J.}\ \bibnamefont {Wiley}},
      \ and\ \bibinfo {author} {\bibfnamefont {K.~I.}\ \bibnamefont {Winey}},\
      }\bibfield  {title} {\enquote {\bibinfo {title} {Integrating {{Simulations}}
      and {{Experiments To Predict Sheet Resistance}} and {{Optical Transmittance}}
      in {{Nanowire Films}} for {{Transparent Conductors}}},}\ }\href {\doibase
      10.1021/nn403324t} {\bibfield  {journal} {\bibinfo  {journal} {ACS Nano}\
      }\textbf {\bibinfo {volume} {7}},\ \bibinfo {pages} {7654--7663} (\bibinfo
      {year} {2013})}\BibitemShut {NoStop}%
    \bibitem [{\citenamefont {O'Callaghan}\ \emph {et~al.}(2016)\citenamefont
      {O'Callaghan}, \citenamefont {da~Rocha}, \citenamefont {Manning},
      \citenamefont {Boland},\ and\ \citenamefont
      {Ferreira}}]{ocallaghan_effective_2016}%
      \BibitemOpen
      \bibfield  {author} {\bibinfo {author} {\bibfnamefont {C.}~\bibnamefont
      {O'Callaghan}}, \bibinfo {author} {\bibfnamefont {C.~G.}\ \bibnamefont
      {da~Rocha}}, \bibinfo {author} {\bibfnamefont {H.~G.}\ \bibnamefont
      {Manning}}, \bibinfo {author} {\bibfnamefont {J.~J.}\ \bibnamefont {Boland}},
      \ and\ \bibinfo {author} {\bibfnamefont {M.~S.}\ \bibnamefont {Ferreira}},\
      }\bibfield  {title} {\enquote {\bibinfo {title} {Effective medium theory for
      the conductivity of disordered metallic nanowire networks},}\ }\href
      {\doibase 10.1039/C6CP05187A} {\bibfield  {journal} {\bibinfo  {journal}
      {Phys. Chem. Chem. Phys.}\ }\textbf {\bibinfo {volume} {18}},\ \bibinfo
      {pages} {27564--27571} (\bibinfo {year} {2016})}\BibitemShut {NoStop}%
    \bibitem [{\citenamefont {He}\ \emph {et~al.}(2018)\citenamefont {He},
      \citenamefont {Xu}, \citenamefont {Qiu}, \citenamefont {He},\ and\
      \citenamefont {Zhou}}]{he_conductivity_2018}%
      \BibitemOpen
      \bibfield  {author} {\bibinfo {author} {\bibfnamefont {S.}~\bibnamefont
      {He}}, \bibinfo {author} {\bibfnamefont {X.}~\bibnamefont {Xu}}, \bibinfo
      {author} {\bibfnamefont {X.}~\bibnamefont {Qiu}}, \bibinfo {author}
      {\bibfnamefont {Y.}~\bibnamefont {He}}, \ and\ \bibinfo {author}
      {\bibfnamefont {C.}~\bibnamefont {Zhou}},\ }\bibfield  {title} {\enquote
      {\bibinfo {title} {Conductivity of two-dimensional disordered nanowire
      networks: Dependence on length-ratio of conducting paths to all nanowires},}\
      }\href {\doibase 10.1063/1.5045176} {\bibfield  {journal} {\bibinfo
      {journal} {J. Appl. Phys.}\ }\textbf {\bibinfo {volume} {124}},\ \bibinfo
      {pages} {054302} (\bibinfo {year} {2018})}\BibitemShut {NoStop}%
    \bibitem [{\citenamefont {Forr{\'o}}\ \emph {et~al.}(2018)\citenamefont
      {Forr{\'o}}, \citenamefont {Demk{\'o}}, \citenamefont {Weydert},
      \citenamefont {V{\"o}r{\"o}s},\ and\ \citenamefont
      {Tybrandt}}]{forro_predictive_2018}%
      \BibitemOpen
      \bibfield  {author} {\bibinfo {author} {\bibfnamefont {C.}~\bibnamefont
      {Forr{\'o}}}, \bibinfo {author} {\bibfnamefont {L.}~\bibnamefont
      {Demk{\'o}}}, \bibinfo {author} {\bibfnamefont {S.}~\bibnamefont {Weydert}},
      \bibinfo {author} {\bibfnamefont {J.}~\bibnamefont {V{\"o}r{\"o}s}}, \ and\
      \bibinfo {author} {\bibfnamefont {K.}~\bibnamefont {Tybrandt}},\ }\bibfield
      {title} {\enquote {\bibinfo {title} {Predictive {{Model}} for the
      {{Electrical Transport}} within {{Nanowire Networks}}},}\ }\href {\doibase
      10.1021/acsnano.8b05406} {\bibfield  {journal} {\bibinfo  {journal} {ACS
      Nano}\ }\textbf {\bibinfo {volume} {12}},\ \bibinfo {pages} {11080--11087}
      (\bibinfo {year} {2018})}\BibitemShut {NoStop}%
    \bibitem [{\citenamefont {Kumar}\ and\ \citenamefont
      {Kulkarni}(2016)}]{kumar_evaluating_2016}%
      \BibitemOpen
      \bibfield  {author} {\bibinfo {author} {\bibfnamefont {A.}~\bibnamefont
      {Kumar}}\ and\ \bibinfo {author} {\bibfnamefont {G.~U.}\ \bibnamefont
      {Kulkarni}},\ }\bibfield  {title} {\enquote {\bibinfo {title} {Evaluating
      conducting network based transparent electrodes from geometrical
      considerations},}\ }\href {\doibase 10.1063/1.4939280} {\bibfield  {journal}
      {\bibinfo  {journal} {J. Appl. Phys.}\ }\textbf {\bibinfo {volume} {119}},\
      \bibinfo {pages} {015102} (\bibinfo {year} {2016})}\BibitemShut {NoStop}%
    \bibitem [{\citenamefont {Xian}\ \emph {et~al.}(2017)\citenamefont {Xian},
      \citenamefont {Han}, \citenamefont {Li}, \citenamefont {Yang}, \citenamefont
      {Wu}, \citenamefont {Lu}, \citenamefont {Gao}, \citenamefont {Zeng},
      \citenamefont {Wang}, \citenamefont {Bai}, \citenamefont {Naughton},
      \citenamefont {Zhou}, \citenamefont {Liu}, \citenamefont {Kempa},\ and\
      \citenamefont {Gao}}]{xian_practical_2017}%
      \BibitemOpen
      \bibfield  {author} {\bibinfo {author} {\bibfnamefont {Z.}~\bibnamefont
      {Xian}}, \bibinfo {author} {\bibfnamefont {B.}~\bibnamefont {Han}}, \bibinfo
      {author} {\bibfnamefont {S.}~\bibnamefont {Li}}, \bibinfo {author}
      {\bibfnamefont {C.}~\bibnamefont {Yang}}, \bibinfo {author} {\bibfnamefont
      {S.}~\bibnamefont {Wu}}, \bibinfo {author} {\bibfnamefont {X.}~\bibnamefont
      {Lu}}, \bibinfo {author} {\bibfnamefont {X.}~\bibnamefont {Gao}}, \bibinfo
      {author} {\bibfnamefont {M.}~\bibnamefont {Zeng}}, \bibinfo {author}
      {\bibfnamefont {Q.}~\bibnamefont {Wang}}, \bibinfo {author} {\bibfnamefont
      {P.}~\bibnamefont {Bai}}, \bibinfo {author} {\bibfnamefont {M.~J.}\
      \bibnamefont {Naughton}}, \bibinfo {author} {\bibfnamefont {G.}~\bibnamefont
      {Zhou}}, \bibinfo {author} {\bibfnamefont {J.-M.}\ \bibnamefont {Liu}},
      \bibinfo {author} {\bibfnamefont {K.}~\bibnamefont {Kempa}}, \ and\ \bibinfo
      {author} {\bibfnamefont {J.}~\bibnamefont {Gao}},\ }\bibfield  {title}
      {\enquote {\bibinfo {title} {A {{Practical ITO Replacement Strategy}}:
      Sputtering-{{Free Processing}} of a {{Metallic Nanonetwork}}},}\ }\href
      {\doibase 10.1002/admt.201700061} {\bibfield  {journal} {\bibinfo  {journal}
      {Adv. Mater. Technol.}\ }\textbf {\bibinfo {volume} {2}},\ \bibinfo {pages}
      {1700061} (\bibinfo {year} {2017})}\BibitemShut {NoStop}%
    \bibitem [{\citenamefont {Azani}\ \emph {et~al.}(2019)\citenamefont {Azani},
      \citenamefont {Hassanpour}, \citenamefont {Tarasevich}, \citenamefont
      {Vodolazskaya},\ and\ \citenamefont {Eserkepov}}]{azani_transparent_2019}%
      \BibitemOpen
      \bibfield  {author} {\bibinfo {author} {\bibfnamefont {M.-R.}\ \bibnamefont
      {Azani}}, \bibinfo {author} {\bibfnamefont {A.}~\bibnamefont {Hassanpour}},
      \bibinfo {author} {\bibfnamefont {Y.~Y.}\ \bibnamefont {Tarasevich}},
      \bibinfo {author} {\bibfnamefont {I.~V.}\ \bibnamefont {Vodolazskaya}}, \
      and\ \bibinfo {author} {\bibfnamefont {A.~V.}\ \bibnamefont {Eserkepov}},\
      }\bibfield  {title} {\enquote {\bibinfo {title} {Transparent electrodes with
      nanorings: A computational point of view},}\ }\href {\doibase
      10.1063/1.5099933} {\bibfield  {journal} {\bibinfo  {journal} {J. Appl.
      Phys.}\ }\textbf {\bibinfo {volume} {125}},\ \bibinfo {pages} {234903}
      (\bibinfo {year} {2019})}\BibitemShut {NoStop}%
    \bibitem [{\citenamefont {Akhunzhanov}, \citenamefont {Tarasevich},\ and\
      \citenamefont {Vodolazskaya}(2020)}]{akhunzhanov_circles_2020}%
      \BibitemOpen
      \bibfield  {author} {\bibinfo {author} {\bibfnamefont {R.~K.}\ \bibnamefont
      {Akhunzhanov}}, \bibinfo {author} {\bibfnamefont {Y.~Y.}\ \bibnamefont
      {Tarasevich}}, \ and\ \bibinfo {author} {\bibfnamefont {I.~V.}\ \bibnamefont
      {Vodolazskaya}},\ }\bibfield  {title} {\enquote {\bibinfo {title} {Circles of
      equal radii randomly placed on a plane: Some rigorous results, asymptotic
      behavior, and application to transparent electrodes},}\ }\href {\doibase
      10.1088/1742-5468/ab74cd} {\bibfield  {journal} {\bibinfo  {journal} {J.
      Stat. Mech: Theory Exp.}\ }\textbf {\bibinfo {volume} {2020}},\ \bibinfo
      {pages} {033202} (\bibinfo {year} {2020})}\BibitemShut {NoStop}%
    \bibitem [{\citenamefont {Kim}\ and\ \citenamefont
      {Torquato}(1990)}]{kim_determination_1990-1}%
      \BibitemOpen
      \bibfield  {author} {\bibinfo {author} {\bibfnamefont {I.~C.}\ \bibnamefont
      {Kim}}\ and\ \bibinfo {author} {\bibfnamefont {S.}~\bibnamefont {Torquato}},\
      }\bibfield  {title} {\enquote {\bibinfo {title} {Determination of the
      effective conductivity of heterogeneous media by {{Brownian}} motion
      simulation},}\ }\href {\doibase 10.1063/1.346276} {\bibfield  {journal}
      {\bibinfo  {journal} {J. Appl. Phys.}\ }\textbf {\bibinfo {volume} {68}},\
      \bibinfo {pages} {3892--3903} (\bibinfo {year} {1990})}\BibitemShut {NoStop}%
    \bibitem [{\citenamefont {Hashin}\ and\ \citenamefont
      {Shtrikman}(1962)}]{hashin_variational_1962}%
      \BibitemOpen
      \bibfield  {author} {\bibinfo {author} {\bibfnamefont {Z.}~\bibnamefont
      {Hashin}}\ and\ \bibinfo {author} {\bibfnamefont {S.}~\bibnamefont
      {Shtrikman}},\ }\bibfield  {title} {\enquote {\bibinfo {title} {A
      {{Variational Approach}} to the {{Theory}} of the {{Effective Magnetic
      Permeability}} of {{Multiphase Materials}}},}\ }\href {\doibase
      10.1063/1.1728579} {\bibfield  {journal} {\bibinfo  {journal} {J. Appl.
      Phys.}\ }\textbf {\bibinfo {volume} {33}},\ \bibinfo {pages} {3125--3131}
      (\bibinfo {year} {1962})}\BibitemShut {NoStop}%
    \bibitem [{\citenamefont {Torquato}\ and\ \citenamefont
      {Chen}(2018)}]{torquato_multifunctional_2018}%
      \BibitemOpen
      \bibfield  {author} {\bibinfo {author} {\bibfnamefont {S.}~\bibnamefont
      {Torquato}}\ and\ \bibinfo {author} {\bibfnamefont {D.}~\bibnamefont
      {Chen}},\ }\bibfield  {title} {\enquote {\bibinfo {title} {Multifunctional
      hyperuniform cellular networks: Optimality, anisotropy and disorder},}\
      }\href {\doibase 10.1088/2399-7532/aaca91} {\bibfield  {journal} {\bibinfo
      {journal} {Multifunct. Mater.}\ }\textbf {\bibinfo {volume} {1}},\ \bibinfo
      {pages} {015001} (\bibinfo {year} {2018})}\BibitemShut {NoStop}%
    \bibitem [{\citenamefont {M.~Akinoglu}\ \emph {et~al.}(2013)\citenamefont
      {M.~Akinoglu}, \citenamefont {Sun}, \citenamefont {Gao}, \citenamefont
      {Giersig}, \citenamefont {Ren},\ and\ \citenamefont
      {Kempa}}]{akinoglu_evidence_2013}%
      \BibitemOpen
      \bibfield  {author} {\bibinfo {author} {\bibfnamefont {E.}~\bibnamefont
      {M.~Akinoglu}}, \bibinfo {author} {\bibfnamefont {T.}~\bibnamefont {Sun}},
      \bibinfo {author} {\bibfnamefont {J.}~\bibnamefont {Gao}}, \bibinfo {author}
      {\bibfnamefont {M.}~\bibnamefont {Giersig}}, \bibinfo {author} {\bibfnamefont
      {Z.}~\bibnamefont {Ren}}, \ and\ \bibinfo {author} {\bibfnamefont
      {K.}~\bibnamefont {Kempa}},\ }\bibfield  {title} {\enquote {\bibinfo {title}
      {Evidence for critical scaling of plasmonic modes at the percolation
      threshold in metallic nanostructures},}\ }\href {\doibase 10.1063/1.4826535}
      {\bibfield  {journal} {\bibinfo  {journal} {Appl. Phys. Lett.}\ }\textbf
      {\bibinfo {volume} {103}},\ \bibinfo {pages} {171106} (\bibinfo {year}
      {2013})}\BibitemShut {NoStop}%
    \bibitem [{\citenamefont {Qiu}\ \emph {et~al.}(2017)\citenamefont {Qiu},
      \citenamefont {Akinoglu}, \citenamefont {Luo}, \citenamefont {Giersig},
      \citenamefont {Liang}, \citenamefont {Ning},\ and\ \citenamefont
      {Zhi}}]{qiu_shape_2017}%
      \BibitemOpen
      \bibfield  {author} {\bibinfo {author} {\bibfnamefont {T.}~\bibnamefont
      {Qiu}}, \bibinfo {author} {\bibfnamefont {E.~M.}\ \bibnamefont {Akinoglu}},
      \bibinfo {author} {\bibfnamefont {B.}~\bibnamefont {Luo}}, \bibinfo {author}
      {\bibfnamefont {M.}~\bibnamefont {Giersig}}, \bibinfo {author} {\bibfnamefont
      {M.}~\bibnamefont {Liang}}, \bibinfo {author} {\bibfnamefont
      {J.}~\bibnamefont {Ning}}, \ and\ \bibinfo {author} {\bibfnamefont
      {L.}~\bibnamefont {Zhi}},\ }\bibfield  {title} {\enquote {\bibinfo {title}
      {Shape {{Control}} of {{Periodic Metallic Nanostructures}} for {{Transparent
      Conductive Films}}},}\ }\href {\doibase 10.1002/ppsc.201600262} {\bibfield
      {journal} {\bibinfo  {journal} {Part. Part. Syst. Char.}\ }\textbf {\bibinfo
      {volume} {34}},\ \bibinfo {pages} {1600262} (\bibinfo {year}
      {2017})}\BibitemShut {NoStop}%
    \bibitem [{\citenamefont {Kim}\ and\ \citenamefont
      {Torquato}(2019)}]{kim_methodology_2019}%
      \BibitemOpen
      \bibfield  {author} {\bibinfo {author} {\bibfnamefont {J.}~\bibnamefont
      {Kim}}\ and\ \bibinfo {author} {\bibfnamefont {S.}~\bibnamefont {Torquato}},\
      }\bibfield  {title} {\enquote {\bibinfo {title} {Methodology to construct
      large realizations of perfectly hyperuniform disordered packings},}\ }\href
      {\doibase 10.1103/PhysRevE.99.052141} {\bibfield  {journal} {\bibinfo
      {journal} {Phys. Rev. E}\ }\textbf {\bibinfo {volume} {99}},\ \bibinfo
      {pages} {052141} (\bibinfo {year} {2019})}\BibitemShut {NoStop}%
    \bibitem [{\citenamefont {Torquato}(2018)}]{Torquato2018_review}%
      \BibitemOpen
      \bibfield  {author} {\bibinfo {author} {\bibfnamefont {S.}~\bibnamefont
      {Torquato}},\ }\bibfield  {title} {\enquote {\bibinfo {title} {Hyperuniform
      {S}tates of {M}atter},}\ }\href {\doibase 10.1016/j.physrep.2018.03.001}
      {\bibfield  {journal} {\bibinfo  {journal} {Phys. Rep.}\ }\textbf {\bibinfo
      {volume} {745}},\ \bibinfo {pages} {1 -- 95} (\bibinfo {year}
      {2018})}\BibitemShut {NoStop}%
    \bibitem [{\citenamefont {Torquato}(2002)}]{torquato_random_2002}%
      \BibitemOpen
      \bibfield  {author} {\bibinfo {author} {\bibfnamefont {S.}~\bibnamefont
      {Torquato}},\ }\href@noop {} {\emph {\bibinfo {title} {Random {{Heterogeneous
      Materials}}: {{Microstructure}} and {{Macroscopic Properties}}}}},\ \bibinfo
      {series} {{\emph{Interdisciplinary }}{{{\emph{Applied Mathematics}}}}}\
      No.~\bibinfo {number} {16}\ (\bibinfo  {publisher} {{Springer Science \&
      Business Media}},\ \bibinfo {address} {{New York}},\ \bibinfo {year}
      {2002})\BibitemShut {NoStop}%
    \bibitem [{\citenamefont {Perrins}, \citenamefont {McKenzie},\ and\
      \citenamefont {McPhedran}(1979)}]{perrins_transport_1979}%
      \BibitemOpen
      \bibfield  {author} {\bibinfo {author} {\bibfnamefont {W.~T.}\ \bibnamefont
      {Perrins}}, \bibinfo {author} {\bibfnamefont {D.~R.}\ \bibnamefont
      {McKenzie}}, \ and\ \bibinfo {author} {\bibfnamefont {R.~C.}\ \bibnamefont
      {McPhedran}},\ }\bibfield  {title} {\enquote {\bibinfo {title} {Transport
      properties of regular arrays of cylinders},}\ }\href {\doibase
      10.1098/rspa.1979.0160} {\bibfield  {journal} {\bibinfo  {journal} {Proc. R.
      Soc. Lond. A}\ }\textbf {\bibinfo {volume} {369}},\ \bibinfo {pages}
      {207--225} (\bibinfo {year} {1979})}\BibitemShut {NoStop}%
    \bibitem [{\citenamefont {Han}\ \emph {et~al.}(2016)\citenamefont {Han},
      \citenamefont {Peng}, \citenamefont {Li}, \citenamefont {Rong}, \citenamefont
      {Ding}, \citenamefont {Akinoglu}, \citenamefont {Wu}, \citenamefont {Wang},
      \citenamefont {Lu}, \citenamefont {Wang}, \citenamefont {Zhou}, \citenamefont
      {Liu}, \citenamefont {Ren}, \citenamefont {Giersig}, \citenamefont
      {Herczynski}, \citenamefont {Kempa},\ and\ \citenamefont
      {Gao}}]{han_optimization_2016}%
      \BibitemOpen
      \bibfield  {author} {\bibinfo {author} {\bibfnamefont {B.}~\bibnamefont
      {Han}}, \bibinfo {author} {\bibfnamefont {Q.}~\bibnamefont {Peng}}, \bibinfo
      {author} {\bibfnamefont {R.}~\bibnamefont {Li}}, \bibinfo {author}
      {\bibfnamefont {Q.}~\bibnamefont {Rong}}, \bibinfo {author} {\bibfnamefont
      {Y.}~\bibnamefont {Ding}}, \bibinfo {author} {\bibfnamefont {E.~M.}\
      \bibnamefont {Akinoglu}}, \bibinfo {author} {\bibfnamefont {X.}~\bibnamefont
      {Wu}}, \bibinfo {author} {\bibfnamefont {X.}~\bibnamefont {Wang}}, \bibinfo
      {author} {\bibfnamefont {X.}~\bibnamefont {Lu}}, \bibinfo {author}
      {\bibfnamefont {Q.}~\bibnamefont {Wang}}, \bibinfo {author} {\bibfnamefont
      {G.}~\bibnamefont {Zhou}}, \bibinfo {author} {\bibfnamefont {J.-M.}\
      \bibnamefont {Liu}}, \bibinfo {author} {\bibfnamefont {Z.}~\bibnamefont
      {Ren}}, \bibinfo {author} {\bibfnamefont {M.}~\bibnamefont {Giersig}},
      \bibinfo {author} {\bibfnamefont {A.}~\bibnamefont {Herczynski}}, \bibinfo
      {author} {\bibfnamefont {K.}~\bibnamefont {Kempa}}, \ and\ \bibinfo {author}
      {\bibfnamefont {J.}~\bibnamefont {Gao}},\ }\bibfield  {title} {\enquote
      {\bibinfo {title} {Optimization of hierarchical structure and
      nanoscale-enabled plasmonic refraction for window electrodes in
      photovoltaics},}\ }\href {\doibase 10.1038/ncomms12825} {\bibfield  {journal}
      {\bibinfo  {journal} {Nat. Commun.}\ }\textbf {\bibinfo {volume} {7}},\
      \bibinfo {pages} {12825} (\bibinfo {year} {2016})}\BibitemShut {NoStop}%
    \bibitem [{\citenamefont {Oskooi}\ \emph {et~al.}(2010)\citenamefont {Oskooi},
      \citenamefont {Roundy}, \citenamefont {Ibanescu}, \citenamefont {Bermel},
      \citenamefont {Joannopoulos},\ and\ \citenamefont
      {Johnson}}]{oskooi_meep_2010}%
      \BibitemOpen
      \bibfield  {author} {\bibinfo {author} {\bibfnamefont {A.~F.}\ \bibnamefont
      {Oskooi}}, \bibinfo {author} {\bibfnamefont {D.}~\bibnamefont {Roundy}},
      \bibinfo {author} {\bibfnamefont {M.}~\bibnamefont {Ibanescu}}, \bibinfo
      {author} {\bibfnamefont {P.}~\bibnamefont {Bermel}}, \bibinfo {author}
      {\bibfnamefont {J.~D.}\ \bibnamefont {Joannopoulos}}, \ and\ \bibinfo
      {author} {\bibfnamefont {S.~G.}\ \bibnamefont {Johnson}},\ }\bibfield
      {title} {\enquote {\bibinfo {title} {Meep: A flexible free-software package
      for electromagnetic simulations by the {{FDTD}} method},}\ }\href {\doibase
      10.1016/j.cpc.2009.11.008} {\bibfield  {journal} {\bibinfo  {journal}
      {Comput. Phys. Commun.}\ }\textbf {\bibinfo {volume} {181}},\ \bibinfo
      {pages} {687--702} (\bibinfo {year} {2010})}\BibitemShut {NoStop}%
    \bibitem [{\citenamefont {Taflove}, \citenamefont {Johnson},\ and\
      \citenamefont {Oskooi}(2013)}]{taflove_advances_2013}%
      \BibitemOpen
      \bibfield  {author} {\bibinfo {author} {\bibfnamefont {A.}~\bibnamefont
      {Taflove}}, \bibinfo {author} {\bibfnamefont {S.~G.}\ \bibnamefont
      {Johnson}}, \ and\ \bibinfo {author} {\bibfnamefont {A.}~\bibnamefont
      {Oskooi}},\ }\href@noop {} {\emph {\bibinfo {title} {Advances in {{FDTD
      Computational Electrodynamics}}: Photonics and {{Nanotechnology}}}}}\
      (\bibinfo  {publisher} {{Artech House}},\ \bibinfo {address} {{Boston}},\
      \bibinfo {year} {2013})\BibitemShut {NoStop}%
    \bibitem [{\citenamefont {Ghaemi}\ \emph {et~al.}(1998)\citenamefont {Ghaemi},
      \citenamefont {Thio}, \citenamefont {Grupp}, \citenamefont {Ebbesen},\ and\
      \citenamefont {Lezec}}]{ghaemi1998surface}%
      \BibitemOpen
      \bibfield  {author} {\bibinfo {author} {\bibfnamefont {H.}~\bibnamefont
      {Ghaemi}}, \bibinfo {author} {\bibfnamefont {T.}~\bibnamefont {Thio}},
      \bibinfo {author} {\bibfnamefont {D.~e.~a.}\ \bibnamefont {Grupp}}, \bibinfo
      {author} {\bibfnamefont {T.~W.}\ \bibnamefont {Ebbesen}}, \ and\ \bibinfo
      {author} {\bibfnamefont {H.}~\bibnamefont {Lezec}},\ }\bibfield  {title}
      {\enquote {\bibinfo {title} {Surface plasmons enhance optical transmission
      through subwavelength holes},}\ }\href@noop {} {\bibfield  {journal}
      {\bibinfo  {journal} {Phys. Rev. B}\ }\textbf {\bibinfo {volume} {58}},\
      \bibinfo {pages} {6779} (\bibinfo {year} {1998})}\BibitemShut {NoStop}%
    \bibitem [{\citenamefont {Ebbesen}\ \emph {et~al.}(1998)\citenamefont
      {Ebbesen}, \citenamefont {Lezec}, \citenamefont {Ghaemi}, \citenamefont
      {Thio},\ and\ \citenamefont {Wolff}}]{ebbesen1998extraordinary}%
      \BibitemOpen
      \bibfield  {author} {\bibinfo {author} {\bibfnamefont {T.~W.}\ \bibnamefont
      {Ebbesen}}, \bibinfo {author} {\bibfnamefont {H.~J.}\ \bibnamefont {Lezec}},
      \bibinfo {author} {\bibfnamefont {H.}~\bibnamefont {Ghaemi}}, \bibinfo
      {author} {\bibfnamefont {T.}~\bibnamefont {Thio}}, \ and\ \bibinfo {author}
      {\bibfnamefont {P.~A.}\ \bibnamefont {Wolff}},\ }\bibfield  {title} {\enquote
      {\bibinfo {title} {Extraordinary optical transmission through sub-wavelength
      hole arrays},}\ }\href@noop {} {\bibfield  {journal} {\bibinfo  {journal}
      {Nature}\ }\textbf {\bibinfo {volume} {391}},\ \bibinfo {pages} {667--669}
      (\bibinfo {year} {1998})}\BibitemShut {NoStop}%
    \bibitem [{\citenamefont {Park}\ \emph {et~al.}(2008)\citenamefont {Park},
      \citenamefont {Mirin}, \citenamefont {Lassiter}, \citenamefont {Nehl},
      \citenamefont {Halas},\ and\ \citenamefont {Nordlander}}]{park2008optical}%
      \BibitemOpen
      \bibfield  {author} {\bibinfo {author} {\bibfnamefont {T.-H.}\ \bibnamefont
      {Park}}, \bibinfo {author} {\bibfnamefont {N.}~\bibnamefont {Mirin}},
      \bibinfo {author} {\bibfnamefont {J.~B.}\ \bibnamefont {Lassiter}}, \bibinfo
      {author} {\bibfnamefont {C.~L.}\ \bibnamefont {Nehl}}, \bibinfo {author}
      {\bibfnamefont {N.~J.}\ \bibnamefont {Halas}}, \ and\ \bibinfo {author}
      {\bibfnamefont {P.}~\bibnamefont {Nordlander}},\ }\bibfield  {title}
      {\enquote {\bibinfo {title} {Optical properties of a nanosized hole in a thin
      metallic film},}\ }\href@noop {} {\bibfield  {journal} {\bibinfo  {journal}
      {ACS Nano}\ }\textbf {\bibinfo {volume} {2}},\ \bibinfo {pages} {25--32}
      (\bibinfo {year} {2008})}\BibitemShut {NoStop}%
    \bibitem [{\citenamefont {Wu}\ and\ \citenamefont
      {Tassi}(2014)}]{wu2014broadband}%
      \BibitemOpen
      \bibfield  {author} {\bibinfo {author} {\bibfnamefont {W.}~\bibnamefont
      {Wu}}\ and\ \bibinfo {author} {\bibfnamefont {N.~G.}\ \bibnamefont {Tassi}},\
      }\bibfield  {title} {\enquote {\bibinfo {title} {A broadband plasmonic
      enhanced transparent conductor},}\ }\href@noop {} {\bibfield  {journal}
      {\bibinfo  {journal} {Nanoscale}\ }\textbf {\bibinfo {volume} {6}},\ \bibinfo
      {pages} {7811--7816} (\bibinfo {year} {2014})}\BibitemShut {NoStop}%
    \bibitem [{\citenamefont {Liapis}, \citenamefont {Sfeir},\ and\ \citenamefont
      {Black}(2016)}]{liapis2016plasmonic}%
      \BibitemOpen
      \bibfield  {author} {\bibinfo {author} {\bibfnamefont {A.~C.}\ \bibnamefont
      {Liapis}}, \bibinfo {author} {\bibfnamefont {M.~Y.}\ \bibnamefont {Sfeir}}, \
      and\ \bibinfo {author} {\bibfnamefont {C.~T.}\ \bibnamefont {Black}},\
      }\bibfield  {title} {\enquote {\bibinfo {title} {Plasmonic hole arrays for
      combined photon and electron management},}\ }\href@noop {} {\bibfield
      {journal} {\bibinfo  {journal} {Appl. Phys. Lett.}\ }\textbf {\bibinfo
      {volume} {109}},\ \bibinfo {pages} {201101} (\bibinfo {year}
      {2016})}\BibitemShut {NoStop}%
    \bibitem [{\citenamefont {Sun}\ \emph {et~al.}(2014)\citenamefont {Sun},
      \citenamefont {Fei~Guo}, \citenamefont {Cao}, \citenamefont {Metin~Akinoglu},
      \citenamefont {Wang}, \citenamefont {Giersig}, \citenamefont {Ren},\ and\
      \citenamefont {Kempa}}]{sun2014broadband}%
      \BibitemOpen
      \bibfield  {author} {\bibinfo {author} {\bibfnamefont {T.}~\bibnamefont
      {Sun}}, \bibinfo {author} {\bibfnamefont {C.}~\bibnamefont {Fei~Guo}},
      \bibinfo {author} {\bibfnamefont {F.}~\bibnamefont {Cao}}, \bibinfo {author}
      {\bibfnamefont {E.}~\bibnamefont {Metin~Akinoglu}}, \bibinfo {author}
      {\bibfnamefont {Y.}~\bibnamefont {Wang}}, \bibinfo {author} {\bibfnamefont
      {M.}~\bibnamefont {Giersig}}, \bibinfo {author} {\bibfnamefont
      {Z.}~\bibnamefont {Ren}}, \ and\ \bibinfo {author} {\bibfnamefont
      {K.}~\bibnamefont {Kempa}},\ }\bibfield  {title} {\enquote {\bibinfo {title}
      {A broadband solar absorber with 12 nm thick ultrathin a-si layer by using
      random metallic nanomeshes},}\ }\href@noop {} {\bibfield  {journal} {\bibinfo
       {journal} {Appl. Phys. Lett.}\ }\textbf {\bibinfo {volume} {104}},\ \bibinfo
      {pages} {251119} (\bibinfo {year} {2014})}\BibitemShut {NoStop}%
    \bibitem [{\citenamefont {Yu}\ \emph {et~al.}(2021)\citenamefont {Yu},
      \citenamefont {Qiu}, \citenamefont {Chong}, \citenamefont {Torquato},\ and\
      \citenamefont {Park}}]{yu2021engineered}%
      \BibitemOpen
      \bibfield  {author} {\bibinfo {author} {\bibfnamefont {S.}~\bibnamefont
      {Yu}}, \bibinfo {author} {\bibfnamefont {C.-W.}\ \bibnamefont {Qiu}},
      \bibinfo {author} {\bibfnamefont {Y.}~\bibnamefont {Chong}}, \bibinfo
      {author} {\bibfnamefont {S.}~\bibnamefont {Torquato}}, \ and\ \bibinfo
      {author} {\bibfnamefont {N.}~\bibnamefont {Park}},\ }\bibfield  {title}
      {\enquote {\bibinfo {title} {Engineered disorder in photonics},}\ }\href@noop
      {} {\bibfield  {journal} {\bibinfo  {journal} {Nat. Rev. Mater.}\ }\textbf
      {\bibinfo {volume} {6}},\ \bibinfo {pages} {226--243} (\bibinfo {year}
      {2021})}\BibitemShut {NoStop}%
    \bibitem [{\citenamefont {Kadulkar}\ \emph {et~al.}(2022)\citenamefont
      {Kadulkar}, \citenamefont {Sherman}, \citenamefont {Ganesan},\ and\
      \citenamefont {Truskett}}]{kadulkar2022machine}%
      \BibitemOpen
      \bibfield  {author} {\bibinfo {author} {\bibfnamefont {S.}~\bibnamefont
      {Kadulkar}}, \bibinfo {author} {\bibfnamefont {Z.~M.}\ \bibnamefont
      {Sherman}}, \bibinfo {author} {\bibfnamefont {V.}~\bibnamefont {Ganesan}}, \
      and\ \bibinfo {author} {\bibfnamefont {T.~M.}\ \bibnamefont {Truskett}},\
      }\bibfield  {title} {\enquote {\bibinfo {title} {Machine learning–assisted
      design of material properties},}\ }\href {\doibase
      10.1146/annurev-chembioeng-092220-024340} {\bibfield  {journal} {\bibinfo
      {journal} {Annu. Rev. Chem. Biomol. Eng.}\ }\textbf {\bibinfo {volume}
      {13}},\ \bibinfo {pages} {DOI:10.1146/annurev--chembioeng--092220--024340}
      (\bibinfo {year} {2022})}\BibitemShut {NoStop}%
    \end{thebibliography}
%

\end{document}